\documentclass[vecphys]{svmult}		

\usepackage{makeidx}         
\usepackage{graphicx}        
\usepackage{epsfig}          
\usepackage{multicol}        
\usepackage[bottom]{footmisc}

\usepackage{color}
\usepackage{amsmath,amssymb}

\DeclareMathOperator{\sn}{sn}
\DeclareMathOperator{\cn}{cn}
\DeclareMathOperator{\dn}{dn}
\DeclareMathOperator{\sech}{sech}

\makeindex             

\begin{document}

\def\jcmindex#1{\index{#1}}
\def\myidxeffect#1{{\bf\large #1}}

\title*{Higher-order field theories: $\phi^6$, $\phi^8$ and beyond} 
\titlerunning{Higher-order field theories: $\phi^6$, $\phi^8$ and beyond}
\author{
Avadh Saxena\inst{1}
\and
Ivan C.\ Christov\inst{2}
\and
Avinash Khare\inst{3}
}

\institute{
Theoretical Division and Center for Nonlinear Studies, Los Alamos National Laboratory, Los Alamos, New Mexico 87545, USA \texttt{avadh@lanl.gov}
\and
School of Mechanical Engineering, Purdue University, West Lafayette, Indiana 47907, USA \texttt{christov@purdue.edu}
\and
Department of Physics, Savitribai Phule Pune University, Pune 411007, India \texttt{khare@physics.unipune.ac.in}
}
\maketitle
\abstract
{
The $\phi^4$ model has been the ``workhorse'' of the classical Ginzburg--Landau phenomenological theory of phase transitions and, furthermore, the foundation for a large amount of the now-classical developments in nonlinear science. However, the $\phi^4$ model, in its usual variant (symmetric double-well potential), can only possess two equilibria. Many complex physical systems possess more than two equilibria and, furthermore, the number of equilibria can change as a system parameter (e.g., the temperature in condensed matter physics) is varied. Thus, ``higher-order field theories'' come into play. This chapter discusses recent developments of higher-order field theories, specifically the $\phi^6$, $\phi^8$ models and beyond. We first establish their context in the Ginzburg--Landau theory of successive phase transitions, including a detailed discussion of the symmetric triple well $\phi^6$ potential and its properties. We also note connections between field theories in high-energy physics (e.g., ``bag models'' of quarks within hadrons) and parametric (deformed) $\phi^6$ models. We briefly mention a few salient points about even-higher-order field theories of the $\phi^8$, $\phi^{10}$, etc.\ varieties, including the existence of kinks with power-law tail asymptotics that give rise to long-range interactions. Finally, we conclude with a set of open problems in the context of higher-order scalar fields theories.
}


\section{Introduction}

The mathematical setting of this chapter is $(1+1)D$ field theories. That is to say, we consider a generic spatiotemporal field, $\phi=\phi(x,t)$ (although, in later sections `$\phi$' maybe be replaced by `$u$' depending on the context), and its concordant governing evolution equation. Within the context of classical neutral scalar field theories, the evolution of $\phi$ is determined by a partial differential equation (PDE) that extremizes the \emph{action} functional (in some appropriate ``natural'' dimensionless units):
\begin{equation}\label{1.1a}
	\mathcal{S}[\phi] = \int dt\int dx \, \left[\frac{1}{2}\phi_t^2 - \frac{1}{2}\phi_x^2 - V(\phi) \right],
\end{equation}
where $x$ and $t$ subscripts henceforth denote partial differentiation. The three terms on the right-hand side of Eq.~\eqref{1.1a} denote, respectively, the kinetic energy of the field, the (negative of the) potential energy within the field, and $V$ is some to-be-specified term quantifying the field's \emph{self-interaction}. Specifically, we call $V$ ``the potential'' of the field theory, and it sets the dynamics of the field. As is convention, we term a field theory with a polynomial potential $V$ of degree $n$ as a ``$\phi^n$ field theory.'' The potential is constructed, derived, modeled or conjectured on the basis of the physical behavior of the system under consideration. We refer the reader to \cite{rr,MantonSut,vach} for more detailed textbook introductions to the subject, including physical examples of various field theories in high-energy theoretical physics.

\jcmindex{\myidxeffect{F}!Field theory}

The Euler--Lagrange equation extremizing the action $\mathcal{S}$ from Eq.~\eqref{1.1a} is easily found to be the nonlinear wave (often referred to as Klein--Gordon-type) equation
\begin{equation}\label{eq:nl_wave_eq}
	\phi_{tt} - \phi_{xx} + V'(\phi) = 0\,,
\end{equation}
where a prime denotes differentiation with respect to the argument of the function, here $\phi$. 
Note that, on the basis of the Lorentz invariance of Eq.~\eqref{eq:nl_wave_eq}, in this chapter we are only concerned with its static solutions, i.e., $\phi_t=0$ and $\phi=\phi(x)$ only, with traveling solutions being obtainable from the latter by a boost transformation. Hence, we are interested in how the solutions of the ordinary differential equation
\begin{equation}\label{1.1b}
	\phi_{xx} = V'(\phi)
\end{equation}
are affected by the choice of $V$. Specifically, we will (mostly) discuss ``simple'' polynomials of even degree that possess $\phi \mapsto -\phi$ symmetry (thus endowing the field theory with reflectional, or $Z_2$, symmetry) as choices for the potential.

\jcmindex{\myidxeffect{K}!Klein--Gordon equation}

In this chapter, we discuss the kink (i.e., domain wall) solutions of Eq.~\eqref{1.1b} and their context in the hierarchy of various higher-order field theories, where by ``higher-order'' we either mean that the potential $V$ is a polynomial of degree greater than four or is non-polynomial. For brevity and clarity, we will often refer the reader to the encyclopedic study of the eighth, tenth- and twelfth-degree field theories (and their kink solutions in the presence of degenerate minima), provided in \cite{kcs}.


\section{First- and second-order phase transitions: The need for higher order field theory}
\label{sec:need_hoft}

\jcmindex{\myidxeffect{G}!Ginzburg--Landau theory}

The quartic, $\phi^4$, potential is the ``workhorse'' of the Ginzburg--Landau (phenomenological) theory of superconductivity \cite{Landau,GinzburgLandau,Tinkham}, taking $\phi$ as the \emph{order parameter} of the theory (i.e., the macroscopic wave function of the condensed phase). In this context, the third term (i.e., $V(\phi)$) in Eq.~\eqref{1.1a} is interpreted as the Landau free energy density, while the combination of the second and third terms (i.e., $\frac{1}{2}\phi_x^2 + V(\phi)$) is the full Ginzburg--Landau free energy density, which allows for domain walls of non-vanishing width and energy to exist between various phases (corresponding to equilibria, i.e., minima of $V$) in the system. Specifically, a prototypical example of the \emph{continuous} (or, second-order) phase transition can be modeled by the classical $\phi^4$ (double well) field theory. 

\jcmindex{\myidxeffect{D}!Double well potential}
\jcmindex{\myidxeffect{S}!Second-order phase transition}

\begin{figure}
\center
\includegraphics[width=0.5\textwidth]{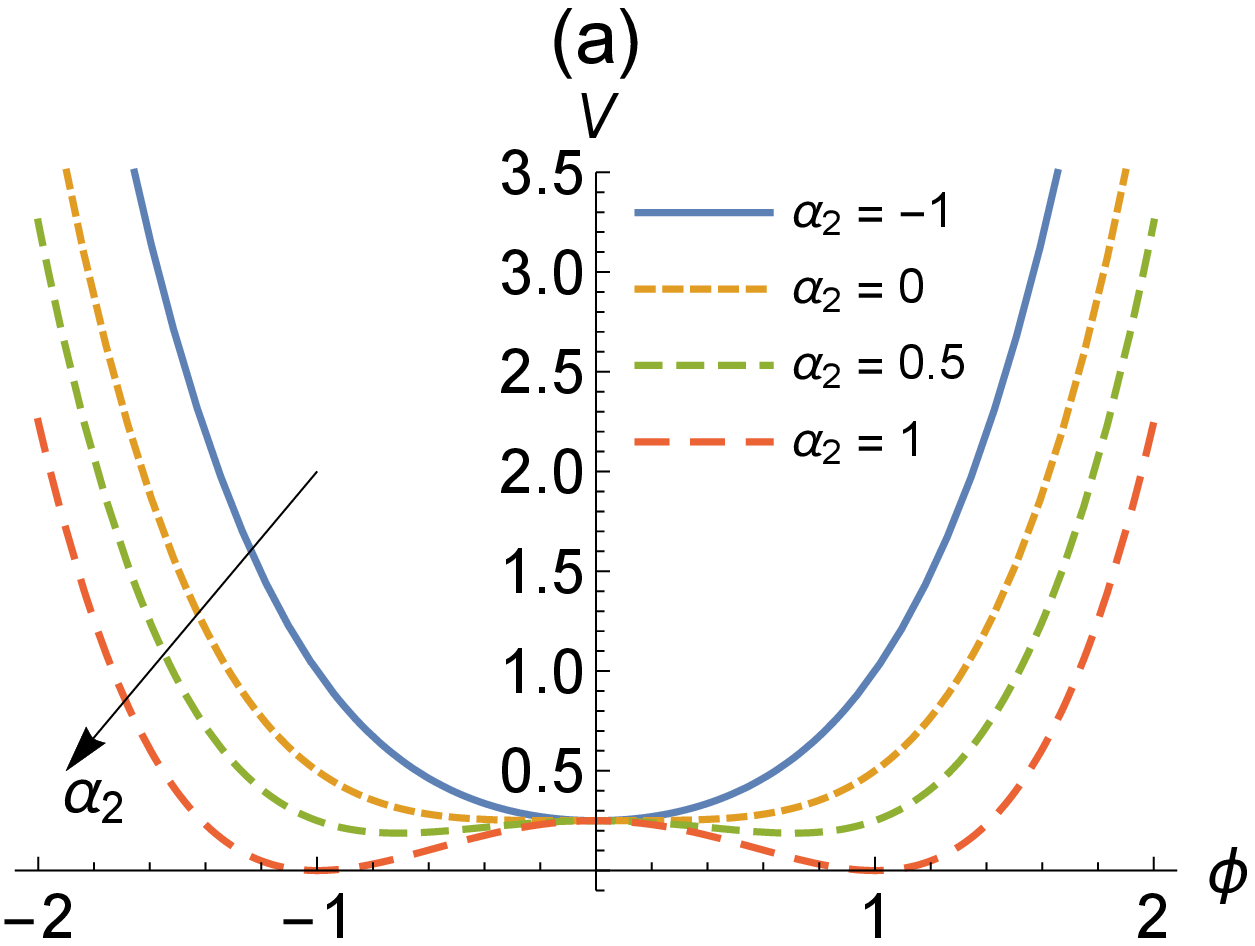}\hfill
\includegraphics[width=0.5\textwidth]{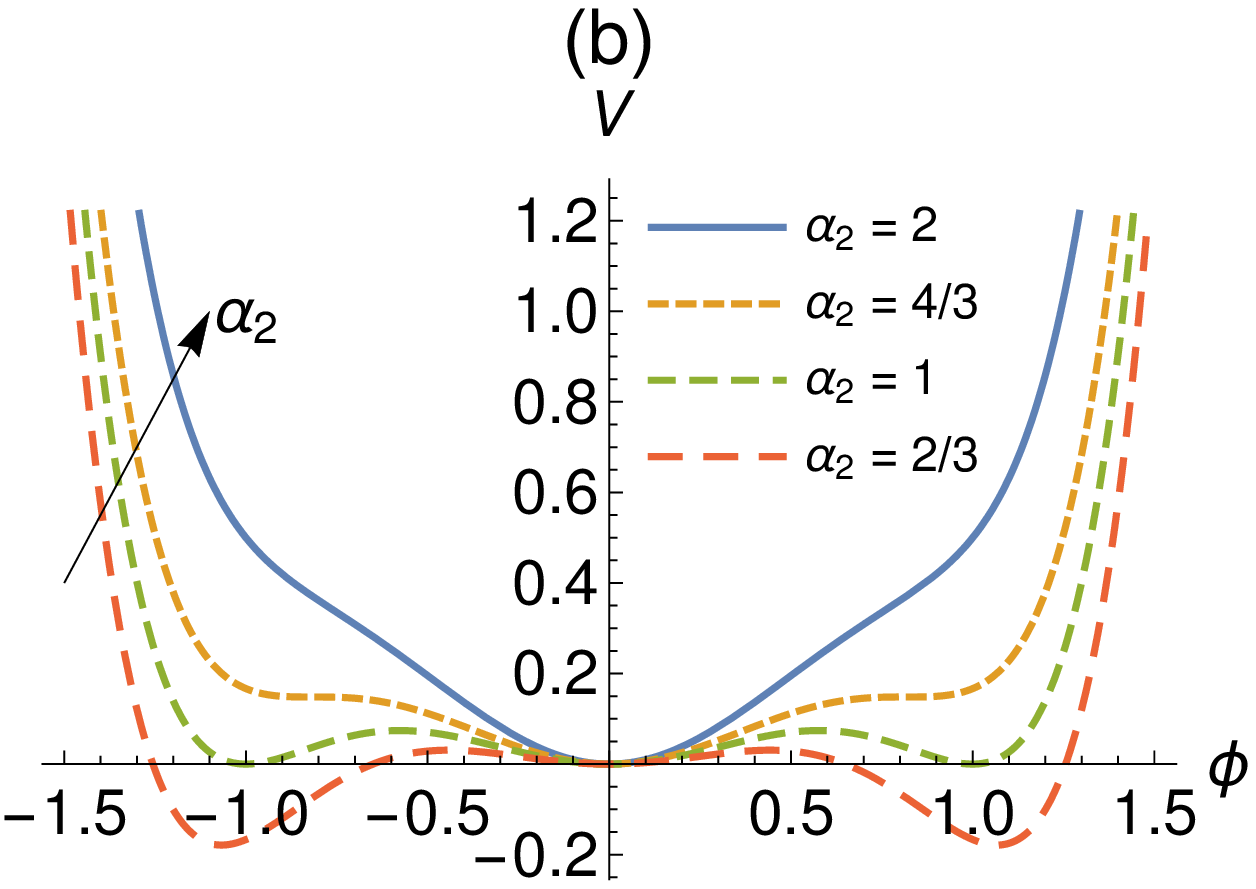}\\[1mm]
\includegraphics[width=0.5\textwidth]{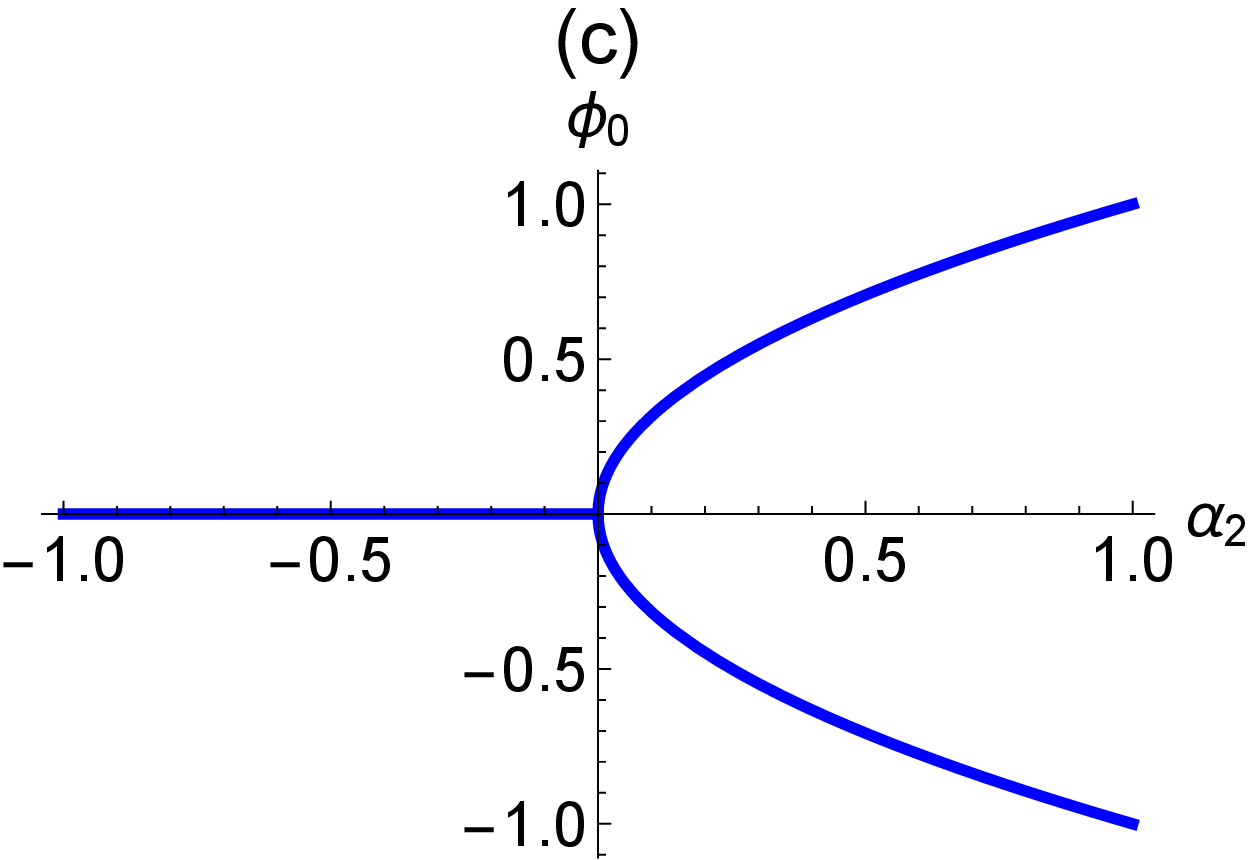}\hfill
\includegraphics[width=0.5\textwidth]{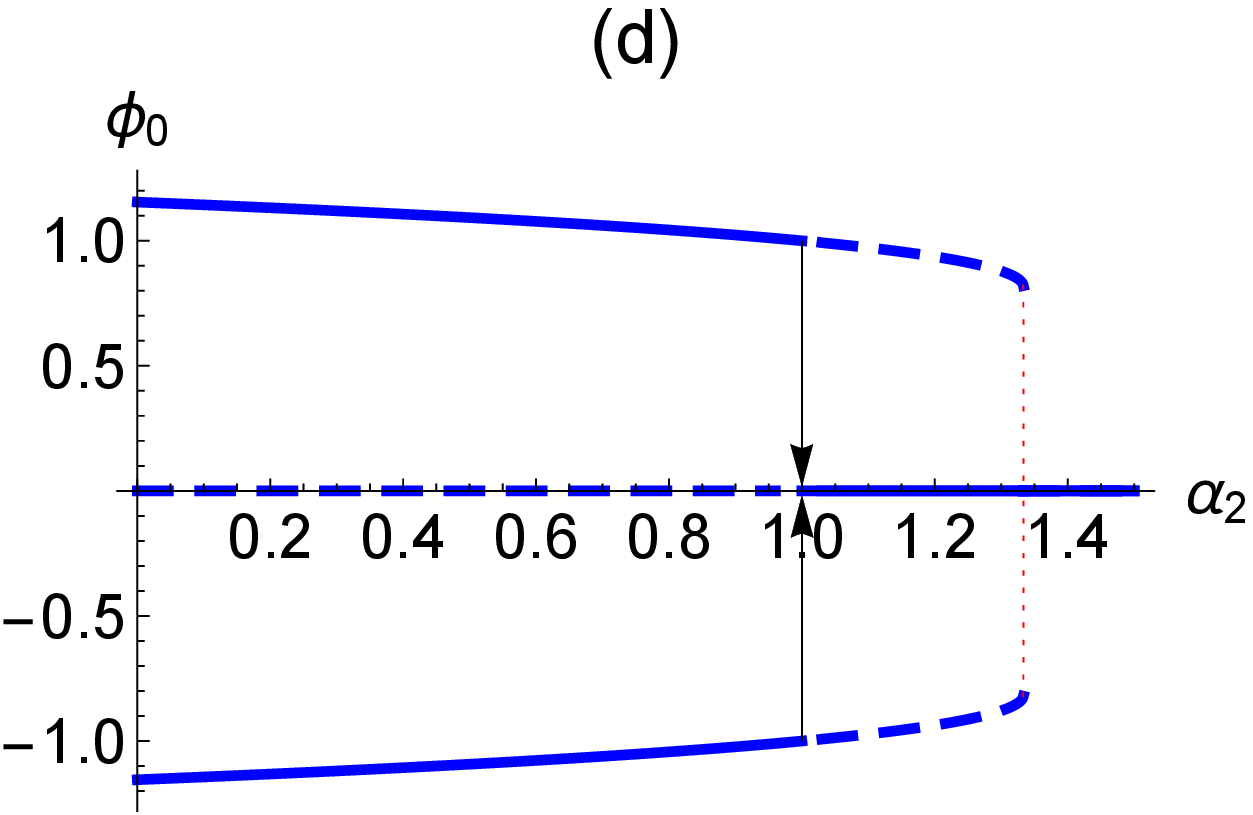}\\[1mm]
\includegraphics[width=0.5\textwidth]{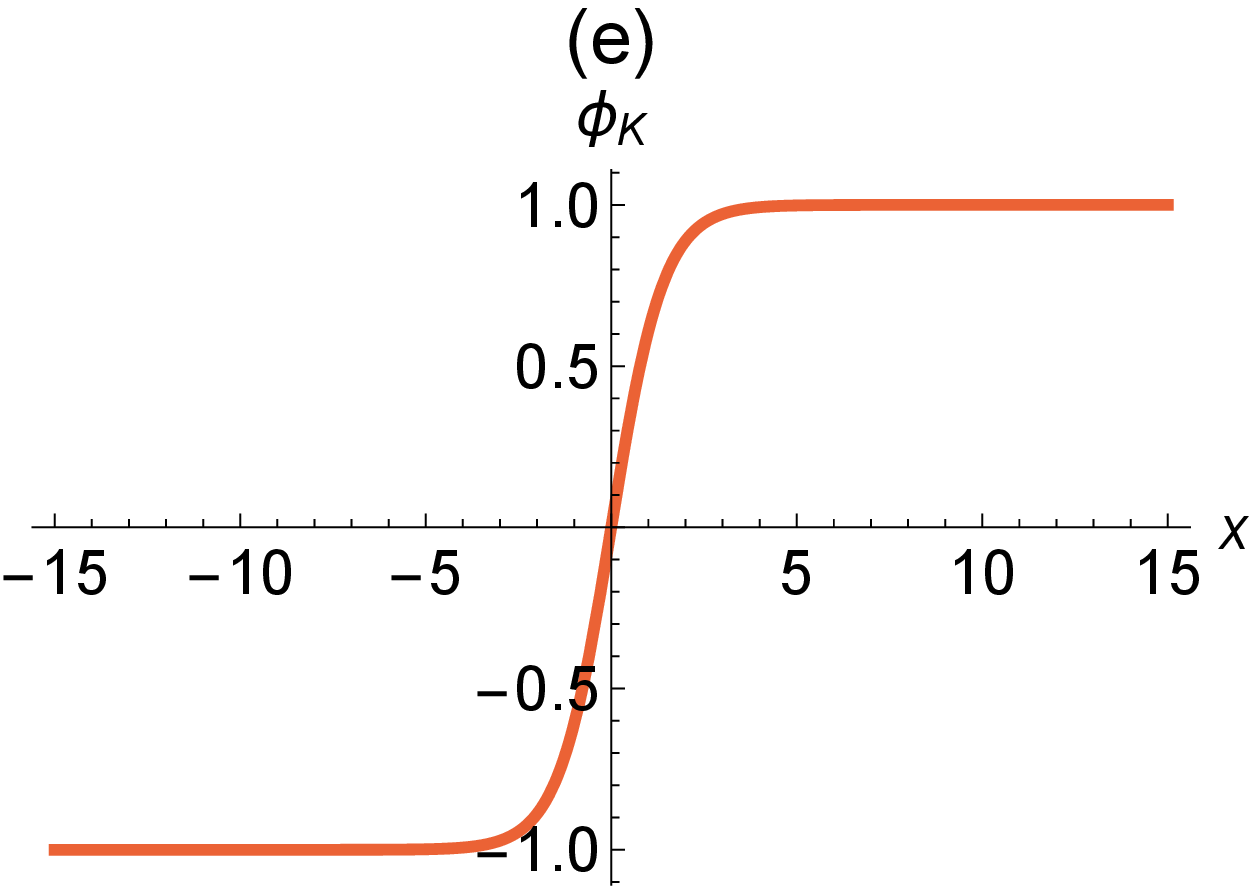}\hfill
\includegraphics[width=0.5\textwidth]{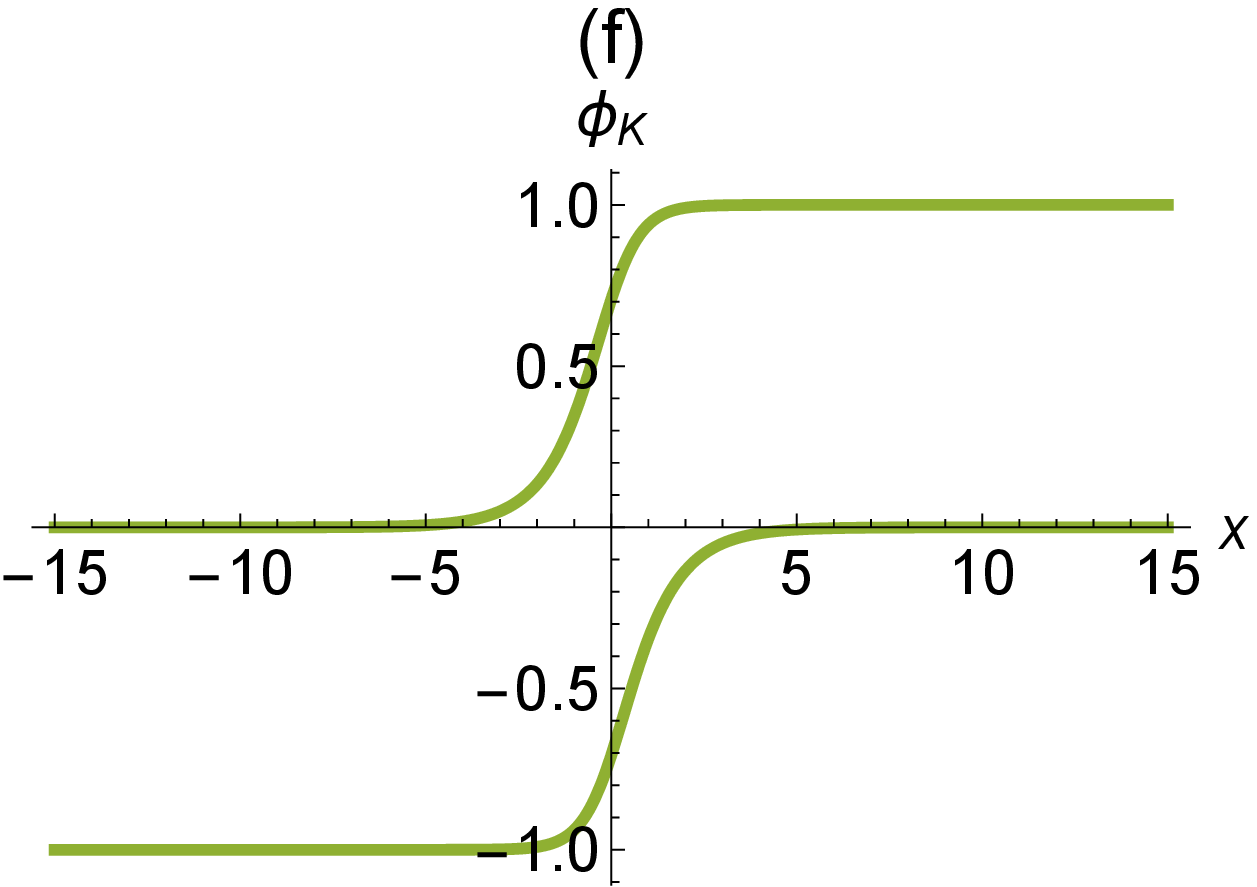}
\caption{Illustrations of (a) a continuous (second-order) phase transition under the $\phi^4$ potential, Eq.~\eqref{eq:V4_pt} and (b) a discontinuous (first-order) phase transition under the $\phi^6$ potential, Eq.~\eqref{eq:V6_pt}. In the literature it is common to shift $V(\phi)\mapsto V(\phi) - V_0(\alpha_2)$ to ensure that the kink solution in Eq.~\eqref{eq:V4_kink} exists for each $\alpha_2$ at which there are two degenerate minima, but we have chosen not to do that here in order to not overcomplicate the picture. Panels (c) and (d) show the respective bifurcations in the minima  $\phi_0$ such that $V'(\phi_0)=0$ and $V''(\phi_0)>0$ as a function of the parameter $\alpha_2$; dashed and solid curves denote the local and global minima, respectively. Panels (e) and (f) show the kink and half-kink solutions, Eqs.~\eqref{eq:V4_kink} and \eqref{eq:V6_hkink}, respectively, at the critical value $\alpha_2=1$ when the potentials in (a) and (b) have two or three degenerate minima.}
\label{fig:46pt_combo}
\end{figure}

To illustrate the second-order phase transition, consider the symmetric quartic (double well) potential
\begin{equation}\label{eq:V4_pt}
V(\phi) = \frac{1}{4}\phi^4-\frac{\alpha_2}{2}\phi^2+\frac{1}{4}\,,
\end{equation}
where $\alpha_2$ is a parameter that might depend on, e.g., the temperature or pressure of the system in condensed matter physics or the mass of a meson in high-energy theoretical physics. As the temperature or pressure of the system changes, so does $\alpha_2$, leading to structural changes of the potential in Eq.~\eqref{eq:V4_pt}, as shown in Fig.~\ref{fig:46pt_combo}(a). Note that at $\alpha_2=1$, Eq.~\eqref{eq:V4_pt} can be rewritten as $V(\phi) = \frac{1}{4}(\phi^2-1)^2$. Specifically, the \emph{global} minima of this potential, i.e., $\phi_0$ such that $V'(\phi_0)=0$ and $V''(\phi_0)>0$, are 
\begin{equation}
\phi_0 = \left\{ \pm \sqrt{\alpha_2} \,\right\} \quad (\alpha_2>0)\,,
\end{equation}
while $\phi_0=0$ is a global maximum. As $\alpha_2\to0^+$, these two minima smoothly coalesce into a single \emph{global} minimum at $\phi_0=0$ ($\alpha_2\le0$), as shown in Fig.~\ref{fig:46pt_combo}(c). This smooth process is characteristic of the continuous, i.e., second-order,  phase transitions, and is the only type of bifurcation of equilibria that a symmetric double well $\phi^4$ potential can exhibit. For $\alpha_2=1$ both degenerate minima of the potential satisfy $V(\phi_0) = V'(\phi_0)=0$, and a domain wall, which solves Eqs.~\eqref{1.1b} and \eqref{eq:V4_pt}, exists connecting the two phases ($\phi_0=-1$ and $\phi_0=+1$):
\begin{equation}\label{eq:V4_kink}
\phi_K(x) = \tanh\left(x/\sqrt{2}\right).
\end{equation}
This well-known domain wall, or \emph{kink}, solution \cite{rr,MantonSut,vach} is illustrated in Fig.~\ref{fig:46pt_combo}(e). 

\jcmindex{\myidxeffect{K}!Kink solution}
\jcmindex{\myidxeffect{D}!Domain wall}

However, in materials science and condensed matter physics, one also observes \emph{discontinuous} (or, first-order) phase transitions, or even \emph{successive} series of first- and/or second-order phase transitions. How can those be modeled? One approach is to add degrees of freedom by increasing the degree of the potential $V$ to greater-than-fourth order \cite{gl}. For example, first-order transitions can be modeled by sextic, $\phi^6$, field theory \cite{Behera,Falk,Falk2}. A triple well potential characteristic of a $\phi^6$ field theory naturally arises as a one-dimensional cross-section across a path of strain space passing through the austenite well and two of the martensite wells of the free energy of a two-phase martensitic material with cubic and tetragonal phases \cite[Sec.~5.5]{Abeyarat}. Although another possibility to model a first-order transition is by way of an \emph{asymmetric} double well $\phi^4$ potential (e.g., a double well potential in an external field) \cite{SanatiSaxenaAJP}, here we restrict ourselves to symmetric potentials only. Then, in order to capture two or more successive transitions, we must go beyond the $\phi^4$ and $\phi^6$ field theories to even higher orders \cite{gl,GufanBook}. Similarly, higher-order field theories arise in the context of high-energy physics, wherein the availability of more than two equilibria leads to more types of mesons \cite{Lohe79,CL}, which is indeed necessary for certain nuclear and particle physics models. 

\jcmindex{\myidxeffect{T}!Triple well potential}
\jcmindex{\myidxeffect{F}!First-order phase transition}

To illustrate the first-order phase transition, consider the sextic (triple well) potential
\begin{equation}\label{eq:V6_pt}
V(\phi) = \frac{1}{2}\phi^6 - \phi^4 + \frac{\alpha_2}{2} \phi^2\,,
\end{equation}
where $\alpha_2$ is again a parameter that might depend on, e.g., the temperature or pressure of the system. Varying $\alpha_2$ leads to structural changes of the potential in Eq.~\eqref{eq:V6_pt}, as shown in Fig.~\ref{fig:46pt_combo}(b). Note that at $\alpha_2=1$, Eq.~\eqref{eq:V6_pt} can be rewritten as $V(\phi) = \frac{1}{2}\phi^2(\phi^2-1)^2$.  Specifically, the minima of this potential are 
\begin{equation}\label{eq:V6_equil}
\phi_0 = 0,\quad \phi_0 = \left\{\pm \frac{\sqrt{\sqrt{4-3 \alpha_2}+2}}{\sqrt{3}} \,\right\} \quad (\alpha_2 < 4/3) \,.
\end{equation}
For $\alpha_2>1$, the two non-zero minima are \emph{local}, while $\phi_0=0$ is the \emph{global} minimum; vice versa for $\alpha_2<1$. At $\alpha_2=1$, the exchange of global minima is \emph{sudden}, i.e., the global minima at $|\phi_0|\ne0$ do not coalesce with the one at $\phi_0=0$ as in the $\phi^4$ example above. This non-smooth process is characteristic of discontinuous, i.e., first-order, phase transitions.  (In Fig.~\ref{fig:46pt_combo}(d), dashed and solid curves denote the local and global minima values of $\phi_0$, respectively.) For $\alpha_2=1$ all three minima of the potential become degenerate and satisfy $V(\phi_0) = V'(\phi_0)=0$, thus domain wall (kink) solutions, which satisfy Eqs.~\eqref{1.1b} and \eqref{eq:V6_pt}, exist connecting a pair of equilibria (either $\phi_0=-1$ and $\phi_0=0$ or $\phi_0=0$ and $\phi_0=+1$):
\begin{equation}\label{eq:V6_hkink}
\phi_K(x) = 
\begin{cases} 
\displaystyle-\frac{1}{\sqrt{\mathrm{e}^{2 x}+1}} = - \sqrt{ \frac{1 - \tanh x}{2} }\,, &\quad \text{half-kink from $-1$ to $0$},\\
\displaystyle\frac{1}{\sqrt{\mathrm{e}^{-2 x}+1}} = \sqrt{ \frac{1 + \tanh x}{2} }\,, &\quad \text{half-kink from $0$ to $+1$},
\end{cases}
\end{equation}
as illustrated in Fig.~\ref{fig:46pt_combo}(f). Equation~\eqref{eq:V6_hkink} represents the well-known $\phi^6$ \emph{half-kink} \cite{Khare79}. Finally, as $\alpha_2\to(4/3)^-$, the two non-zero minima disappear entirely (once again suddenly) leaving a single, global minimum at $\phi_0=0$, as shown in Fig.~\ref{fig:46pt_combo}(d).

\jcmindex{\myidxeffect{H}!Half-kink solution}

Beyond these two introductory examples of a second- and a first-order phase transition, similar reasoning can be applied to show that an octic, $\phi^8$, field theory can model a second-order transition followed by a first-order transition \cite{kcs,R5a,R5b,R6}. Meanwhile, two successive first-order transitions can be modeled by a $\phi^{10}$ field theory \cite{kcs,R7}. But, to describe three successive (first- and/or second-order) transitions one must resort to a $\phi^{12}$ field theory \cite{kcs,R8,R9}. Continuing in the same vein, for four or more successive transitions, a $\phi^{14}$ or higher-order (e.g., $\phi^{4n}$ or $\phi^{4n+2}$ with $n > 3$) field theory must be employed. So far we have implicitly assumed that, as stated at the beginning of the chapter, we deal with neutral scalar (single-component) field theories. Beyond the scope of this chapter but also relevant is the fact that {\it multi-component} $\phi^4$ or $\phi^6$ field theories can also describe successive phase transitions \cite{R10}.

\jcmindex{\myidxeffect{S}!Successive phase transitions}
\jcmindex{\myidxeffect{H}!Higher-order field theory}

Higher-order (specifically, higher than sextic) field theories are also needed to capture \emph{all} symmetry-allowed phases in a transition \cite{R5a,R5b}.  Certain crystals undergo two successive ferroelastic (i.e., strain as the order parameter) or ferroelectric (i.e., electric polarization as the order parameter) first-order transitions \cite{R7}.  In particle physics massless mesons interacting via long-range forces are modeled with the $\phi^8$ field theory \cite{Lohe79}. Additionally, there are examples of isostructural transitions (i.e., the crystal symmetry does not change but the lattice constant changes), which can be described by the $\phi^8$ field theory \cite{R11}. In biophysics, chiral protein crystallization is modeled via a $\phi^{10}$ field theory \cite{R12}. Similarly, the transitions in certain piezoelectric (i.e., stress-induced polarization) materials with perovskite structure are modeled by the $\phi^{12}$ field theory \cite{R8,R9}. 


\section{$\phi^6$ field theory}

As we have just discussed in Sec.~\ref{sec:need_hoft}, the location of the global minimum of a triple well $\phi^6$ potential abruptly (discontinuously) jumps from $\phi_0=0$ to a pair of finite value $|\phi_0|\ne0$ through the phase transition (as $\alpha_2$ goes through $1$ in the example of Eq.~\eqref{eq:V6_pt} above). At the phase transition point, the potential has three global minima. This type of phase transition is ubiquitous in nature: from cosmological transitions in the early Universe \cite{Vilen} to solid-solid transformations from one crystal structure to another \cite{R5a,R5b}. Here, it is relevant to mention the significance of the latter from the thermodynamics point of view (see also \cite[\S5]{Bishop80}): i.e., when the field $\phi(x,t)$ possesses a large number of kinks driven by white noise and balanced by dissipation. At such a discontinuous (first-order) phase transition, described by the $\phi^6$ field theory, we expect that field's self-correlation function will yield finite correlation lengths at the transition temperature, which is associated with latent heat in classical thermodynamics \cite{Reichl}. 

\subsection{Exact kink and periodic solutions, asymptotic kink interaction}
\label{sec:phi6_kink_periodic_etc}

In the case of three degenerate minima, a $\phi^6$ potential can always be factored into the form $\phi^2(\phi^2-a^2)^2$, up to scaling factors, and then the exact domain-wall solutions are the half-kinks in Eq.~\eqref{eq:V6_hkink}. Whether at, above or below the critical temperature ($\alpha_2=1$ for Eq.~\eqref{eq:V6_pt}) at which the system exhibits three stable equilibria, further exact domain-wall solutions exist near the ``wells'' of a $\phi^6$ potential \cite{SanatiSaxenaJPA}.

To illustrate these ideas, let us return to Eq.~\eqref{1.1b}. Multiplying by $\phi_x$ and forming a complete differential, we may integrate both sides to get the \emph{first integral of motion}:
\begin{equation}\label{1.1c}
(\phi_x)^2 = 2[V(\phi) - \mathfrak{C}]\,,
\end{equation}
where $\mathfrak{C}$ is a constant of integration. Assuming $\phi \to \phi_0$ smoothly as $|x|\to\infty$, where $\phi_0$ is a degenerate minimum of $V$ such that $V(\phi_0)=V'(\phi_0)=0$ and $V''(\phi_0)>0$, fixes the integration constant as $\mathfrak{C}=0$. Then, a second integration, taking $V$ to be as in Eq.~\eqref{eq:V6_pt} with $\alpha_2=1$, leads to the solution $\phi_K(x)$ in Eq.~\eqref{eq:V6_hkink}.

\jcmindex{\myidxeffect{K}!Kink lattice solution}

But, what if we do not apply the approach to equilibrium as a boundary condition? Then, what happens when $\mathfrak{C}\ne 0$? To understand this case, note that we may still separate variables in the first-order ODE in Eq.~\eqref{1.1c} to get the implicit relation:
\begin{equation}\label{eq:V6_int}
x = \int \frac{d\phi}{\sqrt{2[V(\phi) - \mathfrak{C}]}}\,,
\end{equation}
where we have yet to specify the limits of integration (hence, no need for a second constant of integration). For clarity, we restrict ourselves to the positive root in Eq.~\eqref{eq:V6_int}. As discussed in \cite{Falk2,SanatiSaxenaJPA}, by picking the integration limits to be consecutive zeros of $V-\mathfrak{C}$, with a maximum of $V$ in between, the right-hand side of Eq.~\eqref{eq:V6_int} becomes an \emph{elliptic integral} \cite{BF54} in the variable $\varphi = \phi^2$, and \emph{Jacobi's elliptic functions} \cite{as} (see also Sec.~\ref{sec:PT_solutions}) can be used to solve for $\phi$. This sets a range for physically admissible choices for $\mathfrak{C}$, namely those between maxima and minima values of $V(\phi)$.

\jcmindex{\myidxeffect{E}!Elliptic integral}

\begin{figure}[t]
\center
\includegraphics[width=\textwidth]{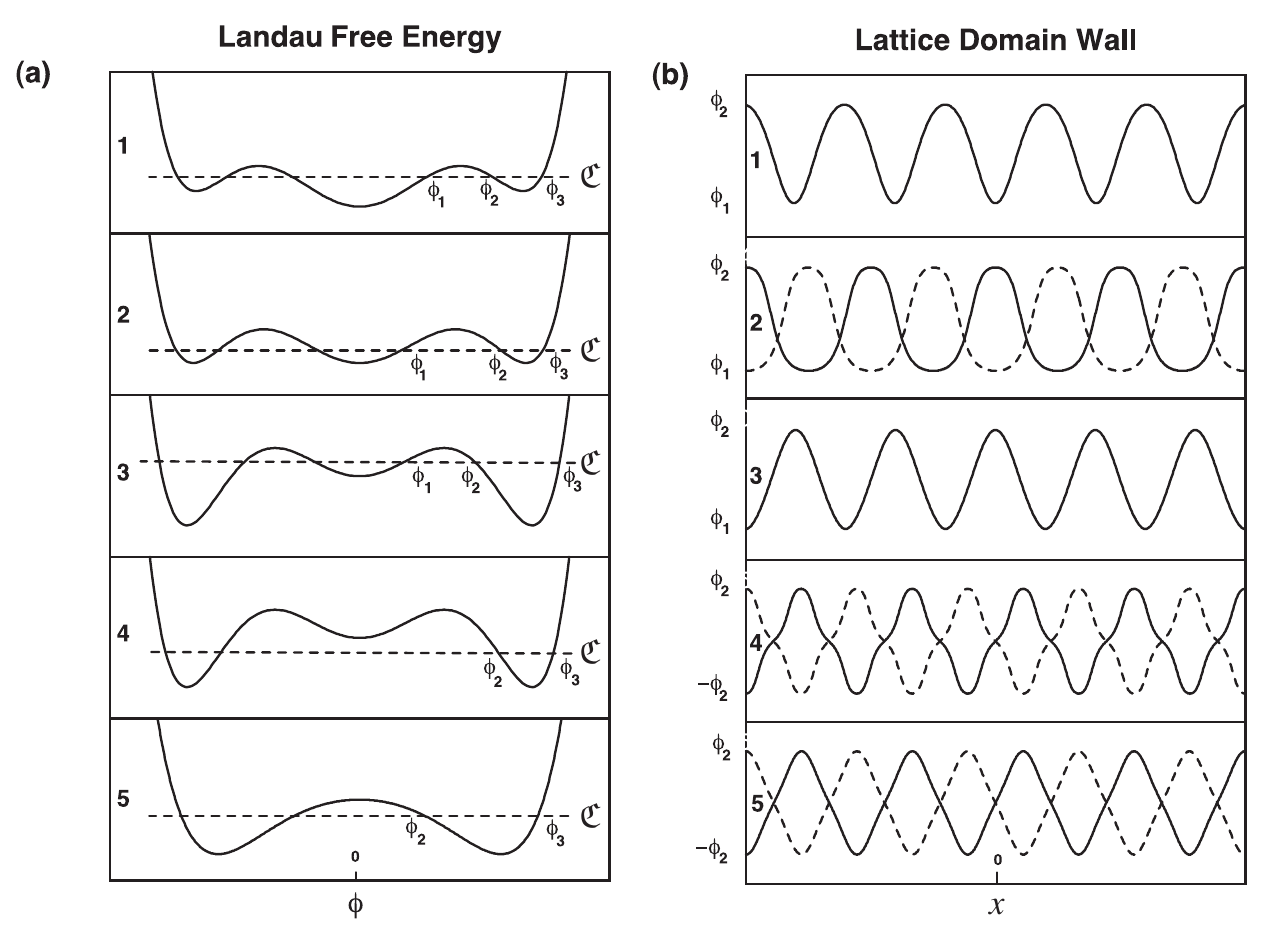}
\caption{(a) A $\phi^6$ potential (i.e., the Landau free energy density of the field theory) at different temperatures (panels 1--5); specifically $V(\phi) = \frac{1}{6}\phi^6 -\frac{1}{4}\phi^4 + \frac{\alpha_2}{2}\phi^2$ for different $\alpha_2$. $\mathfrak{C}$ is the constant of integration in Eq.~\eqref{1.1c}, and $\pm\phi_{1,2,3}$ are the solutions to $V(\phi) = \mathfrak{C}$. (b) Illustrations of the corresponding kink lattices: case 1 [$\frac{3}{16}<\alpha_2<\frac{1}{4}$, $\phi_{KL,1}(x)$], case 2 [$\alpha_2=\frac{3}{16}$, $\phi_{KL,1}(x)$ as dashed and $\phi_{KL,2}(x)$ as solid], case 3 [$0<\alpha_2<\frac{3}{16}$, $\phi_{KL,2}(x)$], cases 4--5 [$\alpha_2<\frac{3}{16}$ or $\alpha_2<0$ \emph{but} $\pm\mathrm{i}\phi_1$ is now a pair of imaginary solutions, $\pm\phi_{KL,3}(x)$ as dashed and solid]. From [M.\ Sanati and A.\ Saxena, ``Half-kink lattice solution of the $\phi^6$ model,'' J.\ Phys.\ A: Math.\ Gen.\ 32, 4311--4320 (1999)] $\copyright$ IOP Publishing. Reproduced with permission.  All rights reserved.}
\label{fig:V6_kink_lattice}
\end{figure}

Figure~\ref{fig:V6_kink_lattice}(b) summarizes visually these so-called \emph{kink lattice} solutions obtained in \cite{SanatiSaxenaJPA} by performing the integration in Eq.~\eqref{eq:V6_int} with $V(\phi) = \frac{1}{6}\phi^6 -\frac{1}{4}\phi^4 + \frac{\alpha_2}{2}\phi^2$ and inverting the expression in terms of the Jacobi elliptic functions $\sn$ and $\dn$:
\begin{subequations}\begin{align}
\phi_{KL,1}(x) &= \frac{\phi_1}{\sqrt{1-A^2\sn^2(\beta x \,|\, m)}}\,,\\
\phi_{KL,2}(x) &= \frac{\phi_2\dn(\beta x,m)}{\sqrt{1-B^2\sn^2(\beta x \,|\, m)}}\,,\\
\phi_{KL,3}(x) &= \frac{\phi_1 C \sn(\bar{\beta} x,\bar{m})}{\sqrt{1-C^2\sn^2(\bar{\beta} x \,|\, \bar{m})}}\,,
\end{align}\label{eq:V6_lattice_kinks}\end{subequations}
where the elliptic moduli $m,\bar{m}\in[0,1]$, and the constants $A$, $B$, $C$, $\beta$ and $\bar{\beta}$ are related to the roots $\phi_{1,2,3}$, satisfying $V(\phi_{1,2,3}) = \mathfrak{C}$ (see Fig.~\ref{fig:V6_kink_lattice}(a)), as
\begin{multline}
A^2 = \frac{\phi_2^2-\phi_1^2}{\phi_2^2},\quad B^2 = \frac{\phi_2^2-\phi_1^2}{\phi_2^3-\phi_1^2},\quad C^2 = \frac{\phi_2^2}{\phi_1^2+\phi_2^2}, \quad \beta^2 = {\tfrac{1}{3}{\phi_2^2(\phi_3^2-\phi_1^2)}}, \\ \bar{\beta} = {\tfrac{1}{3}{\phi_3^2(\phi_2^2+\phi_1^2)}},\quad m = \frac{\phi_3^2(\phi_2^2-\phi_1^2)}{\phi_2^2(\phi_3^2-\phi_1^2)},\quad \bar{m} = \frac{\phi_2^2(\phi_3^3+\phi_1^2)}{\phi_3^2(\phi_2^2+\phi_1^2)}\,.
\end{multline}
The solutions in Eqs.~\eqref{eq:V6_lattice_kinks} are exact and \emph{periodic} with period $4K(m)/\beta$, where $K(m)$ is the complete elliptic integral of the first kind \cite{as}. In cases 2, 4 and 5 (Fig.~\ref{fig:V6_kink_lattice}), these periodic solutions reduce to distinct kink solutions in the limit of $m\to1$ (or $\bar{m}\to 1$, as the case might be).

\jcmindex{\myidxeffect{J}!Jacobi elliptic function}

Finally, the asymptotic force of interaction between $\phi^6$ kinks can be obtained in the usual way via Manton's method \cite{MantonSut,manton_npb} from the exponential tail asymptotics as shown in, e.g., \cite{SanatiSaxenaJPA} (the result is also mentioned in \cite{dorey}).

\jcmindex{\myidxeffect{M}!Manton's method}

\subsection{Linearization about a kink (internal modes) and linearization about an equilibrium (phonon modes)}
\label{sec:V6_linearize}

\jcmindex{\myidxeffect{I}!Internal mode}

Linearizing the field theory about a kink solution, i.e., substituting $\phi(x,t) = \phi_K(x) + \delta \mathrm{e}^{\mathrm{i}\omega_i t}\chi_i(x)+\mathrm{c.c.}$ into Eq.~\eqref{eq:nl_wave_eq} and keeping terms up to $\mathcal{O}(\delta)$, yields a standard Schr\"odinger-type eigenvalue problem \cite{Bishop80}:
\begin{equation}\label{eq:EVP_linearized}
\left[-\frac{d^2}{dx^2} + V''\big(\phi_K(x)\big)\right] \chi_i = \omega_i^2 \chi_i,
\end{equation}
where $\omega_i$ is the temporal oscillation frequency of the $i$th linearization mode, and $\chi_i(x)$ is the eigenfunction giving this mode's spatial structure. The traditional $\phi^4$ symmetric double well's kink, e.g., as in Eq.~\eqref{eq:V4_kink}, possesses a translational mode, $\omega_0=0$, and an internal mode at an isolated eigenvalue at $\omega_1 = \sqrt{3/2}$ (continuous spectrum begins at $\omega = \sqrt{2}$). In fact, it is even possible to write down $\chi_1(x)$ analytically \cite{Bishop80}. Meanwhile, the standard $\phi^6$ symmetric triple well's half-kink, e.g., as in Eq.~\eqref{eq:V6_hkink}, \emph{does not} possess such an internal mode \cite{Khare79,dorey}.

\jcmindex{\myidxeffect{S}!Schr\"odinger-type eigenvalue problem}

However, this issue of whether a single translational mode exists or not, is hardly the whole story in higher-order field theories. As we discuss below, there are $\phi^6$ models with \emph{controllably} many internal modes. Meanwhile, much like the ``classical'' $\phi^4$ and $\phi^6$ pictures, for a $\phi^8$ field theory with four degenerate minima, specifically $V(\phi) = (\phi^2-a^2)^2(\phi^2-b^2)$ with $a,b=(\sqrt{3}\mp1)/2$, which has both full- and half-kink solutions \cite{kcs}, the full-kink has an internal mode ($\omega_1 \approx 1.645$) \cite{GaLeLiconf} while the half-kink does not \cite{GaLeLi}. The possibility of power-law (as opposed to exponential) tails in higher-order field theories adds further complications. The kink of the $\phi^8$ model with two degenerate minima, specifically $V(\phi) = (\phi^2-a^2)^2(\phi^2-b^2)$ with $a=4/5$ and $b=1$, is reported to possess three internal modes ($\omega_1\approx2.068$, $\omega_2\approx3.192$ and $\omega_3\approx3.689$) \cite{GaLeLi}, while kink solutions with power-law tails of other sextic and non-polynomial models possess only the zero mode \cite{Bazeia18}.

\jcmindex{\myidxeffect{P}!Phonon modes}

Meanwhile, phonon modes, i.e., linear excitations about an equilibrium state $\phi_0$, are a simpler matter. Linearizing the field theory about a minimum of $V$ i.e., substituting $\phi(x,t) = \phi_0 + \delta \mathrm{e}^{\mathrm{i}(qx-\omega_q t)}+\mathrm{c.c.}$ into Eq.~\eqref{eq:nl_wave_eq} and keeping terms up to $\mathcal{O}(\delta)$, yields
\begin{equation}\label{eq:phonon_dr}
\omega_q^2 - q^2 = V''(\phi_0) \,,
\end{equation}
for a phonon mode with temporal frequency $\omega_q$ and spatial wave number $q$. For the example triple well $\phi^6$ potential in Eq.~\eqref{eq:V6_pt}, we have $V''(\phi_0) = 15 \phi_0^4 - 12 \phi_0^2 + \alpha_2$. Substituting the equilibria $\phi_0$ from Eq.~\eqref{eq:V6_equil} into the latter gives us
\begin{equation}
V''(\phi_0) = \left\{ \alpha_2,~~ \frac{4}{3} \left(2 \sqrt{4-3 \alpha_2}+4-3 \alpha_2\right),~~ \frac{4}{3} \left(2 \sqrt{4-3 \alpha_2}+4-3 \alpha_2\right) \right\},
\end{equation}
where the second and third values, obviously, hold only for $\alpha_2\le4/3$ (i.e., as long as those minima exist). In particular, at $\alpha_2=1$ for the case of three degenerate minima, we have $V''(\phi_0) = \{ 1,4,4 \}$. Since in all cases we have $V''(\phi_0)\ne0$, then our model $\phi^6$ field theory has well-defined phonon modes along an optical branch (i.e., $\omega_q \not\to 0$ as $|q|\to0$). On the other hand, in certain special cases of higher-than-sixth order field theories (e.g., $\phi^8$), a degeneracy occurs and $V''(\phi_0)=0$, leading to the possibility of \emph{nonlinear phonons}. Nonlinear (or anharmonic) phonons, represent large field excursions of oscillations around the minima of the potential (but do not go over adjacent barriers). Then, in such a case, more terms must be kept in the linearization beyond the vanishing $V''(\phi_0)$ term. 

Finally, we note that by Weyl's theorem, the dispersion relation given by Eq.~\eqref{eq:phonon_dr} describes the continuous spectrum for \emph{both} linearization about a uniform equilibrium and for linearization about a coherent structure such as a kink.

\subsection{Collisional dynamics of $\phi^6$ kinks and multikinks}

{Chapters 2 and 3} discuss the ``classical picture'' of kink--antikink collisions in the $\phi^4$ model as developed/described in the large body of work emanating from \cite{csw,anninos,belova}. In particular, {Chapter 3} discusses some of the recently uncovered twists in this classical picture, as far as the collective-coordinate approach is concerned, and how to resolve them. {Chapter 12} further delves into the notions of fractal structures in the resonance windows and the finer details of their study under the collective-coordinates (variational) approximation. Thus, in this subsection we simply mention one of the more salient aspects of studying kink collisions in higher-order field theories. Specifically, the availability of multiple stable equilibria in the system, which allows for the existence of half-kinks (recall Fig.~\ref{fig:46pt_combo}(f)), opens the possibility of studying collisions between kinks each connecting a \emph{different} pair of equilibria (also called ``topological sectors''). Whereas in the prototypical $\phi^4$ field theory under the potential in Eq.~\eqref{eq:V4_pt} (with $\alpha_2=1$) we only have a kink (given in Eq.~\eqref{eq:V4_kink}) connecting $-1$ to $+1$ or antikink connecting $+1$ to $-1$, in the example $\phi^6$ field theory under the potential in Eq.~\eqref{eq:V6_pt} (with $\alpha_2=1$) we only have \emph{two} half-kinks (given in Eq.~\eqref{eq:V6_hkink}) and their corresponding antikinks. Clearly, this key difference between the $\phi^4$ and $\phi^6$ models gives rise to a potentially far richer phenomenology of kink-kink and kink-antikink collisions.

\begin{figure}[b]
\center
\includegraphics[width=0.5\textwidth]{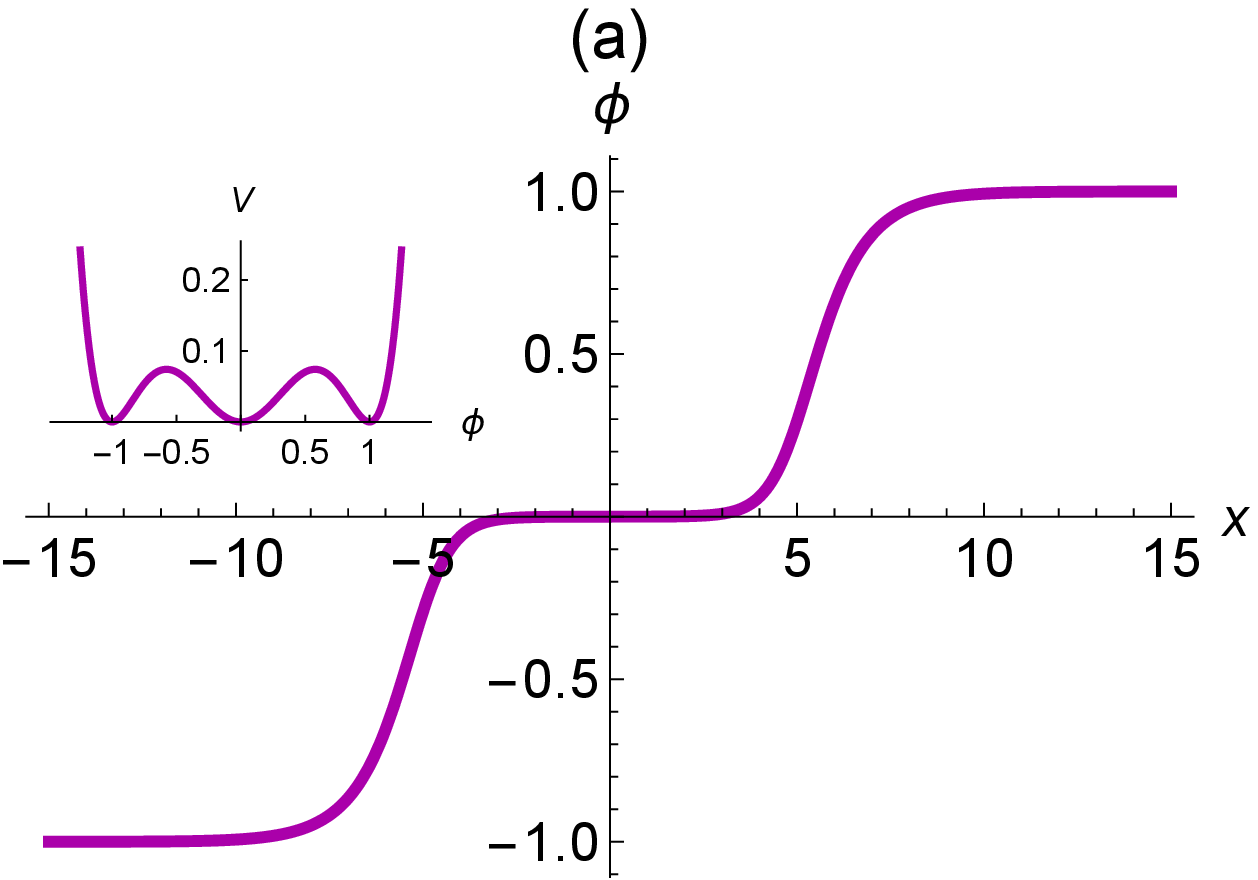}\hfill
\includegraphics[width=0.5\textwidth]{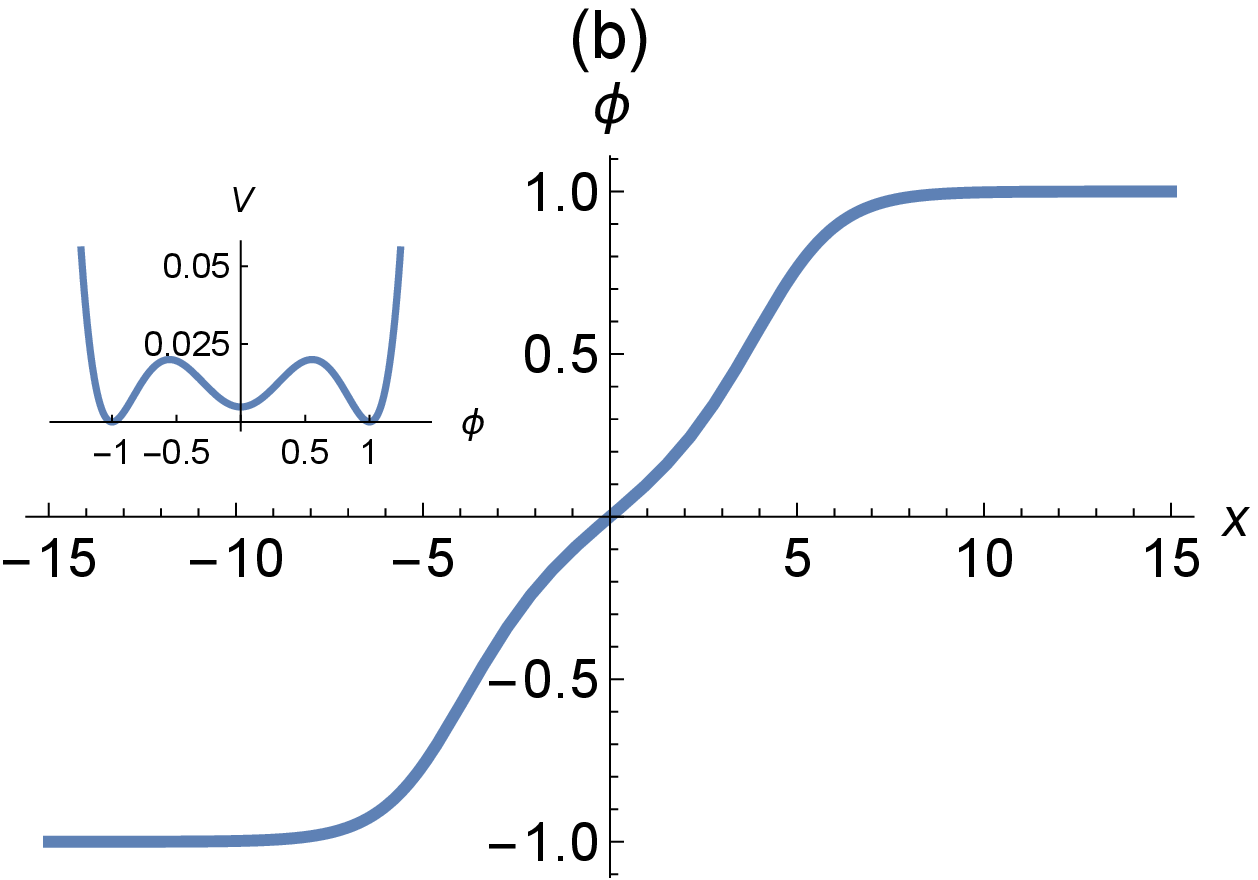}
\caption{Two types of ``staircase'' multikink-type ans\"atze studied in the literature. (a) The example $\phi^6$ field theory, Eq.~\eqref{eq:V6_pt} at $\alpha_2=1$ exhibiting three degenerate minima (see inset), allowing for the superposition of two well-separated half-kinks from Eq.~\eqref{eq:V6_hkink}. (b) The Christ--Lee model, Eq.~\eqref{eq:V_CL} at $\epsilon=0$ exhibiting two degenerate minima (see inset and note the middle, non-degenerate minimum ``lifting off'' from the origin) and a ``bound pair'' exact kink solution given in Eq.~\eqref{eq:CL_bp_kink}.}
\label{fig:CL_G6_combo}
\end{figure}

For example, the collisional dynamics of a ``staircase'' half-kink+half-kink ansatz, which is formed by superimposing the half-kink from $-1$ to $0$ onto the kink from $0$ to $+1$, suitably well separated as shown in Fig.~\ref{fig:CL_G6_combo}(a), were studied by the classical collective coordinate approach in \cite{gani1}, with an updated treatment (resolving certain quantitative discrepancies) given in \cite{weigel1,weigel2}. These types of kink+kink collisions are obviously not possible in the $\phi^4$ model, where one typically studies kink--antikink collisions only. The $\phi^6$ collision phenomenology is, thus, more subtle. Further explorations of multikink configurations, meaning various superpositions of half-kinks in some prescribed arrangements, were presented in \cite{MGSDJ}.

\jcmindex{\myidxeffect{C}!Collective-coordinate approach}
\jcmindex{\myidxeffect{C}!Collisions of kinks}
\jcmindex{\myidxeffect{P}!Parametrically deformed $\phi^6$}

\jcmindex{\myidxeffect{B}!Bound pair kink}

A related possibility in $\phi^6$ field theories is exact kink solutions that look like a ``bound pair'' of kinks (see Fig.~\ref{fig:CL_G6_combo}(b)), similar to the ``staircase'' kink in Fig.~\ref{fig:CL_G6_combo}(a) discussed above. Such kinks can be found in the parametric $\phi^6$ model introduced by Christ and Lee \cite{CL}, specifically an example potential (fixing some of the extra parameters from \cite{CL}) of this form is 
\begin{equation}\label{eq:V_CL}
V(\phi) = \frac{1}{8(1+\epsilon^2)} (\phi^2+\epsilon^2)(\phi^2-1)^2\,.
\end{equation}
The corresponding exact kink solution to Eqs.~\eqref{1.1b} and \eqref{eq:V_CL} (see also \cite{DDKCS}) is
\begin{equation}\label{eq:CL_bp_kink}
\phi_K(x) = \frac{\epsilon\sinh(x/2)}{\sqrt{1 + \epsilon^2 + [\epsilon\sinh(x/2)]^2}}\,.
\end{equation}
Notice that as $\epsilon\to0^+$ or $\epsilon\to\infty$, the potential in Eq.~\eqref{eq:V_CL} takes the form of the prototypical triple well $\phi^6$ potential (i.e., Eq.~\eqref{eq:V6_pt} with $\alpha_2=1$, suitably normalized) or the prototypical double well $\phi^4$ potential (i.e., Eq.~\eqref{eq:V4_pt} with $\alpha_2=1$, suitably normalized), respectively, discussed above. The context of the Christ--Lee model is not condensed matter physics or phase transitions, but rather it was introduced in high-energy theoretical physics as a ``bag" model in which the role of quarks within a hadron is played by the domain wall solutions of the field theory.

For the Christ--Lee model, with potential given by Eq.~\eqref{eq:V_CL}, studying the collisional dynamics of the kink solutions from Eq.~\eqref{eq:CL_bp_kink} yields highly nontrivial results (as compared to the ``classical picture'' of $\phi^4$ kink--antikink collisions). Specifically, as the parameter $\epsilon$ is tuned in the Christ--Lee model, one can \emph{control} the number of internal modes (i.e., non-zero isolated eigenvalues of Eq.~\eqref{eq:EVP_linearized}) of the staircase-like kink. Although it has long been posited \cite{csw} that the internal mode of the kink's linearization (recall Sec.~\ref{sec:V6_linearize}) to a large extent sets the collisional dynamics, recent results using the $\phi^6$ model \cite{dorey} have proposed an additional mechanism unrelated to the internal mode. After the work in \cite{dorey}, it was further shown in \cite{DDKCS} that the resonance window structure exhibits quite counterintuitive behaviors as the number of internal modes in the Christ--Lee model under Eq.~\eqref{eq:V_CL} is tuned. Specifically, this increase in the number of internal modes does \emph{not} lead to more complex resonance structures of ever more multi-bounce windows.  Instead, for a wider range of collision velocities, the staircase-like kinks simply scatter elastically off to infinity. 

\jcmindex{\myidxeffect{M}!Multi-bounce windows}
\jcmindex{\myidxeffect{R}!Resonance windows}

\begin{figure}
\center
\includegraphics[width=0.6\textwidth]{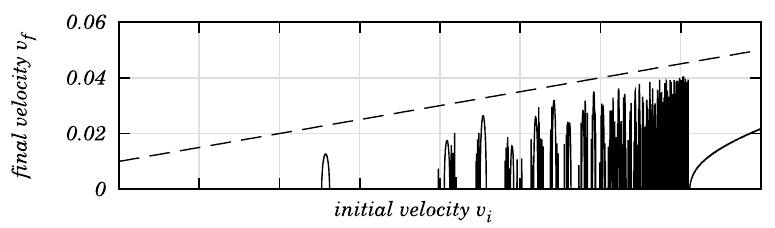}\hfill
\includegraphics[width=0.4\textwidth]{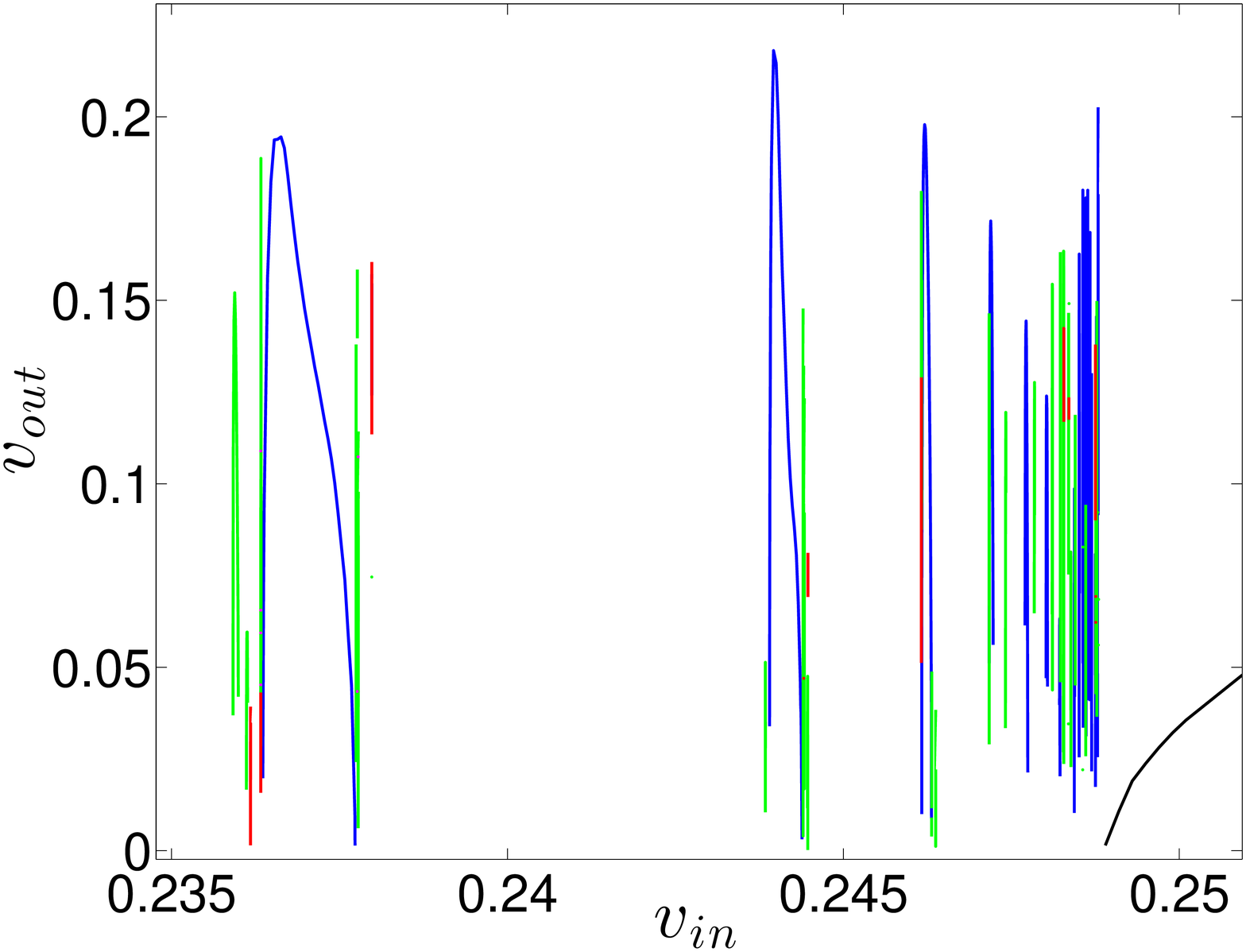}\\
\hspace{4cm}(a)\hfill (b)\hspace{2cm}
\caption{Resonance window maps, based on direct numerical simulation of Eq.~\eqref{eq:nl_wave_eq}, of the final kink velocity ($v_f$ or $v_{out}$) upon a collision of prescribed initial kink velocities ($v_i$ or $v_{in}$). (a) The traditional $\phi^6$ model (i.e., Eq.~\eqref{eq:V6_pt} with $\alpha_2=1$) and no internal modes for the half-kink (i.e., Eq.~\eqref{eq:V6_hkink}). (b) The parametric $\phi^6$ model (i.e., Eq.~\eqref{eq:V_CL} with $\epsilon=0.5$) and four internal modes for the staircase-like kink (i.e., Eq.~\eqref{eq:CL_bp_kink}); colors indicate how many bounces it took to escape (black for one, blue for two, green for three, red for four). Panel (a) is reprinted with permission from [Patrick Dorey, Kieran Mersh, Tomasz Romanczukiewicz, and Yasha Shnir, Physical Review Letters, 107, 091602, 2011] \cite{dorey} $\copyright$ The American Physical Society. Panel (b) is reprinted (without modification) from \cite{DDKCS}, $\copyright~2017$ The Authors of \cite{DDKCS}, under the CC BY 4.0 license.}
\label{fig:res_wind_4_vs_CL}
\end{figure}

\jcmindex{\myidxeffect{I}!Internal mode}

Figure~\ref{fig:res_wind_4_vs_CL} shows a comparison between (a) the ``classical'' $\phi^6$ resonance window (kink with one internal mode) structure of kink collisions and (b) the parametric $\phi^6$ theory under Eq.~\eqref{eq:V_CL} with $\epsilon=0.5$ (i.e., four internal modes of the staircase-like kink). The study of $\phi^4$ kink interactions and resonances is a time-honored subject that has led to elegant demonstrations of Hamiltonian dynamics and even mechanical demonstrations of the two-bounce windows \cite{goodman_chaos}. Following \cite{goodman}, given initially symmetrically located kinks with equal and opposite velocities $v_{in}$, a direct numerical simulation of Eq.~\eqref{eq:nl_wave_eq} is performed, colliding the kinks. If they ``escape'' the collision going off to infinity, the escape velocity $v_{out}$ is recorded and plotted. Clearly, only for some ranges of $v_{in}$ is there a computable $v_{out}$. The ranges in which $v_{out}$ does not exist are termed \emph{resonance} windows in which the kinks continue to bounce back-and-forth forming a bound pair of sorts. Counterintuitively, the structure of these resonance windows in the absence of an internal mode, Fig.~\ref{fig:res_wind_4_vs_CL}(a), is far more complex than in the presence of four internal modes, Fig.~\ref{fig:res_wind_4_vs_CL}(b). We will not delve into this mystery further here because there remain many open problems about kink interactions in $\phi^6$ (and even higher-order) field theories.

\subsection{Statistical mechanics of the $\phi^6$ theory, including QES results}
\label{sec:f6_PDF}

Equation~\eqref{1.1b} subject to the $\phi^6$ potential, e.g., as given in Eq.~\eqref{eq:V6_pt}, represents a highly anharmonic system. Therefore, the number of nonlinear (e.g., soliton and breather) and linear (e.g., phonons) elementary excitations is thermally controlled.  In order to determine the thermal density of these excitations, and their individual contribution to correlation functions (and other thermodynamic quantities such as specific heat and entropy), one must investigate their statistical mechanics.  In one dimension, entropy considerations dictate the presence of kinks. Thus, the interactions between kinks and phonons and possibly other excitations play a crucial role in the overall thermodynamics of the process. This question has been of significant interest in condensed matter physics for the past 40 years \cite[\S5]{Bishop80}. The latter can be studied using a probability density function (PDF), which can be calculated either analytically via the path-integral approximation scheme \cite{SSF72,KS75} or numerically by way of Langevin dynamics \cite{MG90}. In these ways, one can obtain equilibrium properties; and, not just the PDF but also the presence of heterophase fluctuations in the vicinity of a phase transition, the field configuration(s), the average total kink-number density, correlation functions, structure factors, specific heat, internal energy and entropy. 

\jcmindex{\myidxeffect{L}!Langevin dynamics}
\jcmindex{\myidxeffect{K}!Kink field thermodynamics}
\jcmindex{\myidxeffect{T}!Transfer-operator approach}
\jcmindex{\myidxeffect{S}!Statistical mechanics of kinks}

The $\phi^4$ model and its attendant kink field have been extensively studied in the literature using techniques such as the path integral formalism \cite{SSF72,KS75}. As discussed in {Chapter 4}, Langevin dynamics were also developed for computing the thermodynamic quantities of a $\phi^4$ field theory \cite{AH93,Kovner,BHL99,HL00}. For higher-order field theories, on the other hand, not much is known beyond the very preliminary results regarding $\phi^6$ in \cite{Habib}. In general, we expect a much richer phenomenology in terms of the possible kink structures and their interactions, under higher-order field theories.

\jcmindex{\myidxeffect{Q}!Quasi-exactly solvable model}

An important departure of the $\phi^6$ model (and, indeed, all higher-order field theories of the form $\phi^{4n+2}$) from the $\phi^4$ model, is that it leads to a \emph{quasi-exactly solvable} (QES) problem \cite{Leach} for the PDF of the kink field. This result was shown in \cite{Behera,Bruce1980} for $\phi^6$, then some further exact PDFs were obtained for $\phi^{10}$ in \cite{kcs}. Let us illustrate the basic idea of this approach. Via the path-integral (transfer operator) formalism \cite{SSF72,KS75,AH93,Kovner} (see also \cite[Sec.~10.5]{PeyDauBook}), one can reduce the statistical mechanics problem of finding the PDF to solving, once again, a Schr\"odinger-type eigenvalue problem:
\begin{equation}\label{eq:schro_evp}
	\left[-\frac{1}{2\beta^2}\frac{d^2}{d\phi^2} + V(\phi)\right]\Psi_k = E_k\Psi_k,
\end{equation}
where $\beta$ is an inverse temperature, $(\Psi_k, E_k)$ is the sought after eigenpair and $V$ is the model potential. For $V(\phi)$ given in Eq.~\eqref{eq:V6_pt}, Eq.~\eqref{eq:schro_evp} is a well-known QES eigenvalue problem \cite{Ushve}. Specifically, one posits one solution to Eq.~\eqref{eq:schro_evp} (out of the infinite number of possible ones) in the form
\begin{equation}
	\Psi_0(\phi) = \exp\left\{-\frac{1}{2}\phi^2\left(\phi^2 - K\right)\right\} ,
\label{eq:ground_state_phi6}
\end{equation} 
where $E_0$ and $K$ are still to be determined. Upon substituting Eq.~\eqref{eq:ground_state_phi6} for the wavefunction $\Psi_0(\phi)$ and Eq.~\eqref{eq:V6_pt} for $V(\phi)$ into Eq.~\eqref{eq:schro_evp} and requiring that equality hold, one obtains the consistency conditions:
\begin{equation}
\alpha_2 = -\frac{1}{2},\quad K = 2,\quad E_0 = - \frac{1}{4},\quad \beta = 2\,.
\end{equation}
Thus, for the specific $\phi^6$ potential in Eq.~\eqref{eq:V6_pt} with $\alpha_2=-1/2$ and at the precise (inverse) temperature $\beta = 2$, Eq.~\eqref{eq:ground_state_phi6} represents an \emph{exact} ground state PDF (i.e., the wavefunction has no nodes) for the $\phi^6$ field theory, as long as $E_0=-1/4$ and $K=2$. Finally, the PDF for the field is just the normalized squared ground state wave function $\Psi_0^2$ from Eq.~\eqref{eq:ground_state_phi6}.

\jcmindex{\myidxeffect{S}!Schr\"odinger-type eigenvalue problem}

This is but one example for an exact solution, many other ans\"atze that would conceivably lead to further exact PDF solutions are provided in \cite{Ushve}, potentially including excited states. The exactness of these solutions (and the accuracy of the path-integral formalism) can subsequently be verified by Langevin simulations \cite{MG90,Kovner}. Other examples of QES \emph{non-polynomial} field theories that have both exact kink solutions and quasi-exactly solvable thermodynamics are discussed in \cite{SH,KHS,HKS}. Finally, we emphasize once more that the PDF for the $\phi^4$ model \emph{can only be obtained numerically} (or approximately using certain Gaussian fits) \cite{AH93,Kovner}.

\jcmindex{\myidxeffect{L}!Langevin dynamics}


\section{$\phi^8$ field theory}

\subsection{Successive phase transitions}
\label{sec:phi8_spt}

\jcmindex{\myidxeffect{S}!Successive phase transitions}
\jcmindex{\myidxeffect{F}!First-order phase transition}
\jcmindex{\myidxeffect{S}!Second-order phase transition}

\begin{figure}[t]
\center
\includegraphics[width=0.5\textwidth]{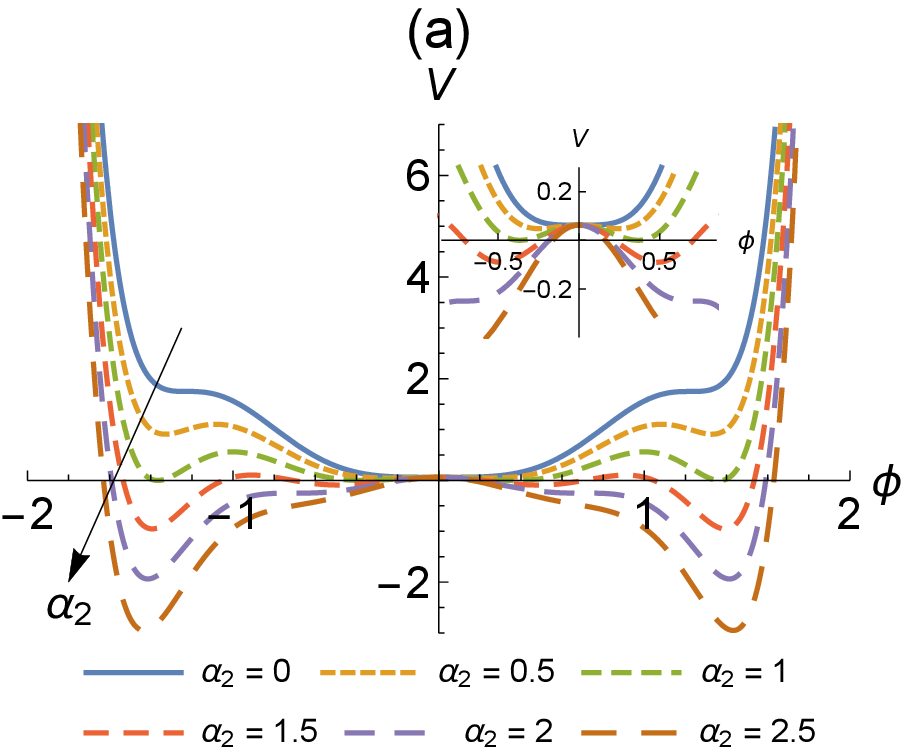}\hfill
\includegraphics[width=0.5\textwidth]{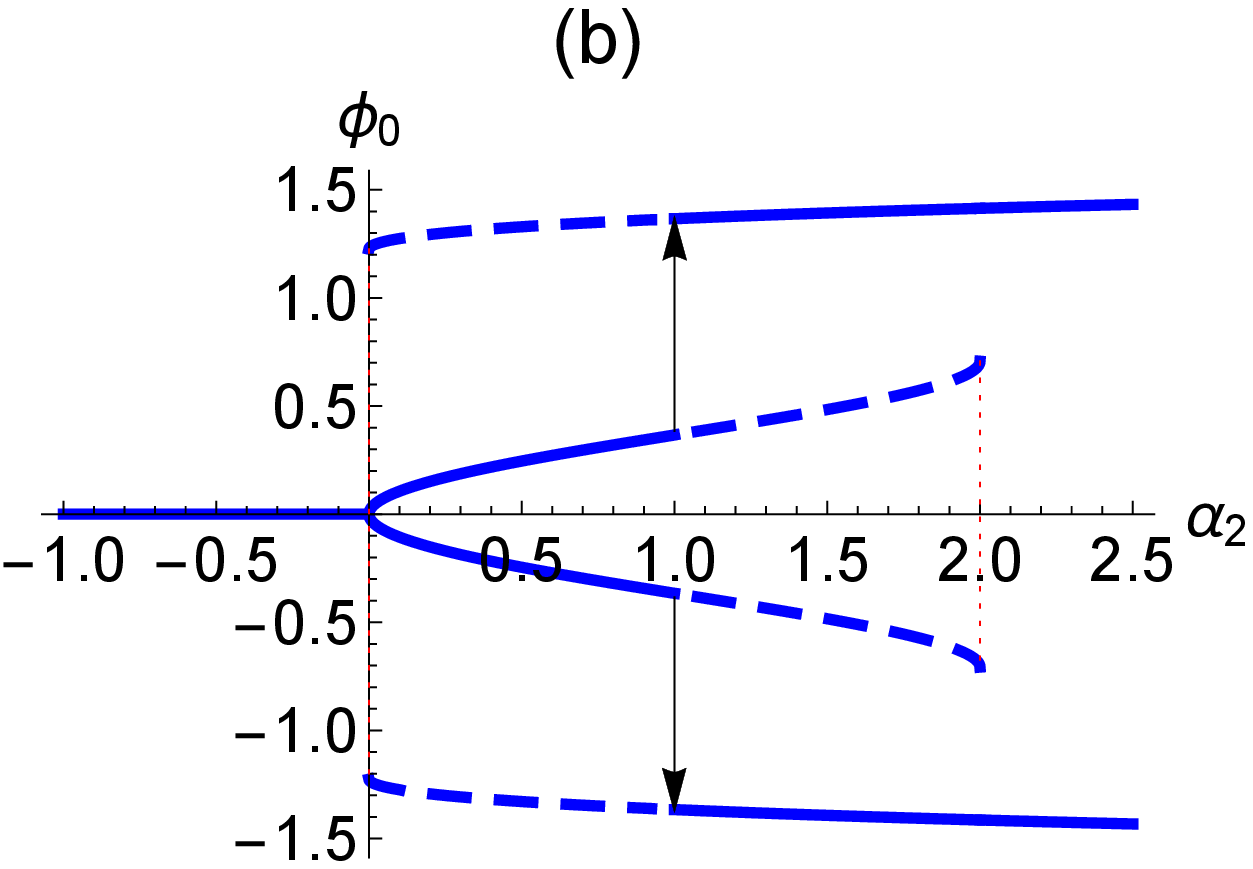}
\caption{(a) Structure of the $\phi^8$ potential in Eq.~\eqref{eq:V8_pt} for different $\alpha_2$ (i.e., values of the coefficient of the quadratic term), showing the various phases and phase transitions in this field theory; inset shows zoom near the origin. (b) Bifurcations in the minima $\phi_0$ such that $V'(\phi_0)=0$ and $V''(\phi_0)>0$ as a function of the parameter $\alpha_2$; dashed and solid curves denote the local and global minima values, respectively.}
\label{fig:phi8_fig1}
\end{figure}

A $\phi^8$ field theory can be used to describe a first-order transition followed by a second-order phase transition. That is to say, as the coefficients of the potential are varied, it is possible to observe coalescence (continuously) and global/local exchange (discontinuously) of minima. A comprehensive discussion is given in \cite[Sec.~II-A]{kcs}. Let us now illustrate, through Fig.~\ref{fig:phi8_fig1} and its discussion, how a succession of a first-order and a second-order phase transition can be described using the octic potential
\begin{equation}\label{eq:V8_pt}
V(\phi) = \phi^{8} - 4\phi^6 + \frac{9}{2} \phi^4 - \alpha_2 \phi^2 + \frac{1}{16},
\end{equation}
where $\alpha_2$ is a free parameter that can be varied to observe the successive phase transitions (e.g., it can be considered a function of the system's temperature). The coefficient of $\phi^8$ in $V$ can be taken to be unity, without loss of generality, by an appropriate rescaling of the $x$-coordinate.  

First, note that, as $\alpha_2\to0^+$, the potential in Eq.~\eqref{eq:V8_pt} has an absolute minimum at $\phi_0=0$ into which two global minima at 
\begin{equation}\label{eq:V8_inner_min}
\phi_{0,\mathrm{inner}} = \pm \frac{1}{2} \sqrt{\mathrm{i} \sqrt{3} \left(r - \frac{1}{r} \right) - r - \frac{1}{r} + 4}\,,\quad r = \sqrt[3]{\sqrt{(\alpha_2-2) \alpha_2}+\alpha_2-1}\,,
\end{equation}
have coalesced. Note that $\phi_{0,\mathrm{inner}}$ are actually real numbers (thus, exist) only for $0\le \alpha_2 \le 2$. Meanwhile the two local minima,
\begin{equation}\label{eq:V8_outer_min}
\phi_{0,\mathrm{outer}} = \pm \sqrt{1 + \frac{r}{2} + \frac{1}{2r}}\qquad (\alpha_2\ge0)\,,
\end{equation}
where $r$ is as given in Eq.~\eqref{eq:V8_inner_min}, have become the inflection points $\phi_{0}=\pm\sqrt{3/2}$ at $\alpha_2=0$. This behavior is analogous to the $\phi^4$ scenario illustrated in Fig.~\ref{fig:46pt_combo}(a). Hence, $\alpha_2=0$ corresponds to a second-order phase transition. 

\jcmindex{\myidxeffect{S}!Second-order phase transition}

Second, note that for $\alpha_2=1$, the potential in Eq.~\eqref{eq:V8_pt} has four (the maximum number of) degenerate global minima, and can be factored into the form $V(\phi) = (\phi^2-a^2)^2(\phi^2-b^2)^2$ with $a=\sqrt{\frac{1}{2} \left(2-\sqrt{3}\right)}$ and $b=\sqrt{\frac{1}{2} \left(2+\sqrt{3}\right)}$. As $\alpha_2$ passes through the value of $1$, the inner pair of minima and the outer pair of minima suddenly exchange their local/global nature.  Hence, $\alpha_2=1$ corresponds to a first-order phase transition temperature of the system. This behavior is analogous to the $\phi^6$ scenario illustrated in Fig.~\ref{fig:46pt_combo}(b).

\jcmindex{\myidxeffect{F}!First-order phase transition} 

Going further, for $\alpha_2=2$, the potential in Eq.~\eqref{eq:V8_pt} has absolute minima at $\phi_{0,\mathrm{outer}}=\pm\sqrt{2}$, a maximum at $\phi_0=0$ and inflection points at $\phi_{0,\mathrm{inner}}=\pm\sqrt{2}/2$. Meanwhile for $1 < \alpha_2 < 2$, the potential in Eq.~\eqref{eq:V8_pt} has global minima at $\phi_{0,\mathrm{outer}}$, as given by Eq.~\eqref{eq:V8_outer_min}, local minima at $\phi_{0,\mathrm{inner}}$, as given by Eq.~\eqref{eq:V8_inner_min}, and three maxima, including one at $\phi_0=0$. Then, for $0 < \alpha_2 < 1$, the situation is reversed and the global minima
are at $\phi_{0,\mathrm{inner}}$, while the local minima are at $\phi_{0,\mathrm{outer}}$; there are still three maxima, including one at $\phi_0 = 0$. For $\alpha_2 < 0$, the example potential has only a single minimum at $\phi_0=0$ and no other extrema.

\subsection{Exact kink solutions: ``The rise of the power-law tails''}
\label{sec:power-law}

\jcmindex{\myidxeffect{P}!Power-law tails of kinks}

A classification and enumeration of kink solutions to $\phi^8$ field theories with degenerate minima can be found in \cite[Sec.~II]{kcs}. First, we note that, given the extra degrees of freedom, an octic potential can have up to four simultaneous degenerate minima, at the first-order phase transition. In this case a kink \emph{and} a half-kink are possible, each with different energy \cite{kcs}. Next, let us now  summarize the most salient feature of these kink solutions: the possibility of \emph{algebraic} (``slow'') decay of the kinks' shapes $\phi_K(x)$ towards the equilibria $\phi_0$ as $|x|\to\infty$, i.e., \emph{power-law} tail asymptotics.

\jcmindex{\myidxeffect{D}!Double well potential}

Consider an octic potential with two degenerate minima (equilibria) at $\phi_0=\pm a$ (i.e., a double well potential), specifically $V(\phi) = \frac{\lambda^2}{2} (\phi^2-a^2)^4$, which has an exact, \emph{implicit} kink solution of Eq.~\eqref{1.1b} \cite[Eq.~(32)]{kcs}:
\begin{equation}\label{7.8}
x(\phi) = \frac{2a\phi}{\gamma_1 (a^2-\phi^2)}+ \frac{1}{\gamma_1}\ln \left(\frac{a+\phi}{a-\phi}\right),
\end{equation}
where $\gamma_1 = 4\lambda a^3$. The implicit relation for $x(\phi)$ in Eq.~\eqref{7.8} can be easily inverted to give $\phi_K(x)$ using, e.g., {\sc Mathematica}. From Eq.~(\ref{7.8}), the asymptotics of the tails of this kink are found to be algebraic (and symmetric) \cite[Eq.~(33)]{kcs}: 
\begin{equation}\label{eq:V8_kink_tail_2dm}
\phi_K(x) \simeq  \mp a\left(1 \pm \displaystyle \frac{1}{\gamma_1 x}\right),\quad x\to \mp \infty\,.
\end{equation}

\jcmindex{\myidxeffect{T}!Triple well potential}

Next, consider an octic potential with three degenerate minima (equilibria) at $\phi_0=0,\pm a$ (i.e., a triple well potential), specifically $V(\phi) = \frac{\lambda^2}{2}\phi^4 (\phi^2-a^2)^2$, which has an exact half-kink solution \cite[Eq.~(23)]{kcs} of Eq.~\eqref{1.1b} given {implicitly} by
\begin{equation}\label{7.2}
x(\phi) = -\frac{2a}{\gamma_2\phi} + \frac{1}{\gamma_2}\ln \left (\frac{a+\phi}{a-\phi} \right ) ,
\end{equation}
where $\gamma_2 = 2\lambda a^3$. (See also \cite[Eq.\ (67)]{Lohe79} but it should be noted that there is a typographical error therein that is evident upon comparison with Eq.~\eqref{7.2}.)
From expanding Eq.~(\ref{7.2}) perturbatively as $|x|\to\infty$, the ``tails'' of the kink can be shown to be of mixed algebraic/exponential (asymmetric) type \cite[Eq.~(24)]{kcs}:
\begin{equation}\label{eq:V8_kink_tail_3dm}
\phi_K(x) \simeq 
a \times \begin{cases}
-\displaystyle \frac{2}{\gamma_2 x},\quad &x\to -\infty\,,\\[3mm]
 1 - \displaystyle {2} e^{- \gamma_2 x -2} ,\quad &x\to +\infty\,.
\end{cases}
\end{equation}

\begin{figure}[t]
\center
\includegraphics[width=0.5\textwidth]{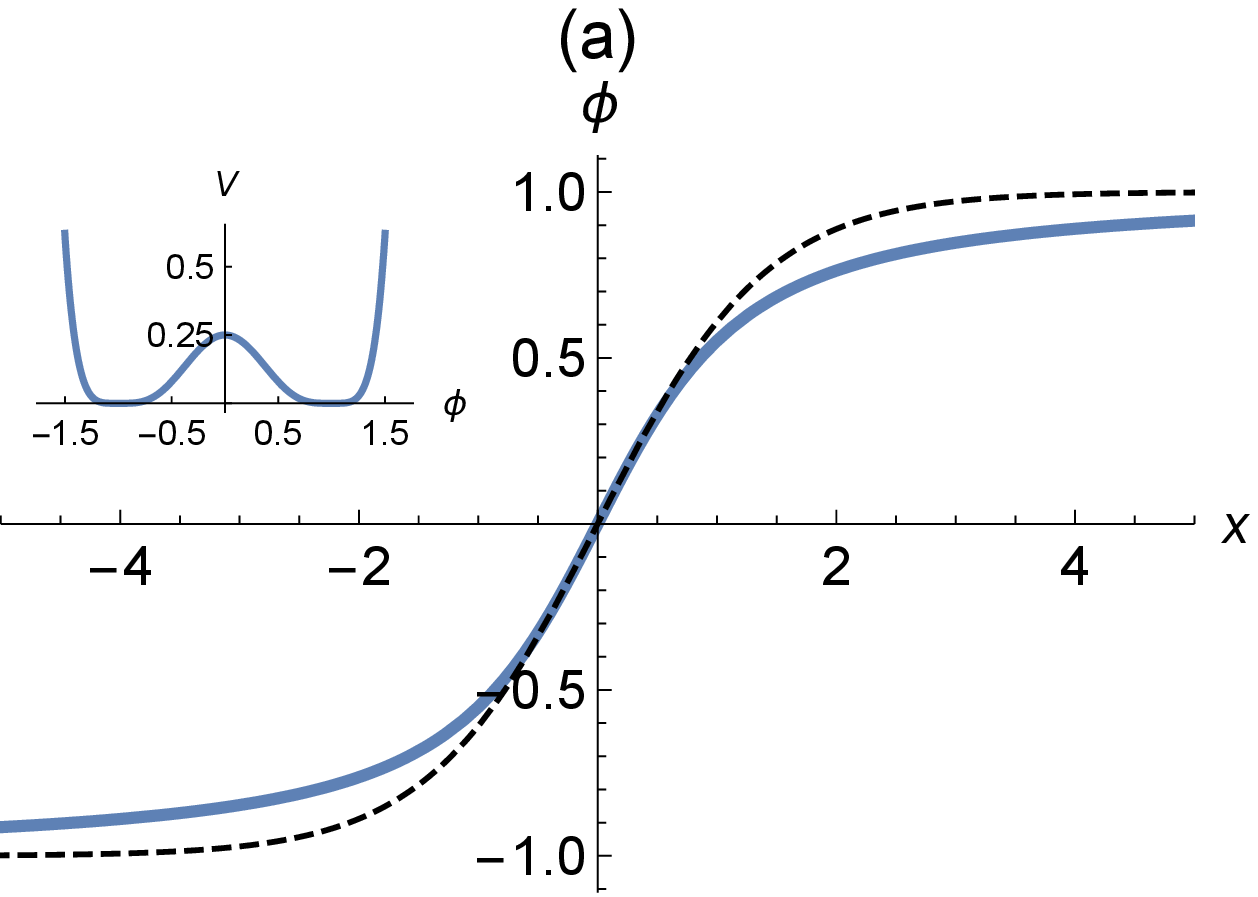}\hfill
\includegraphics[width=0.5\textwidth]{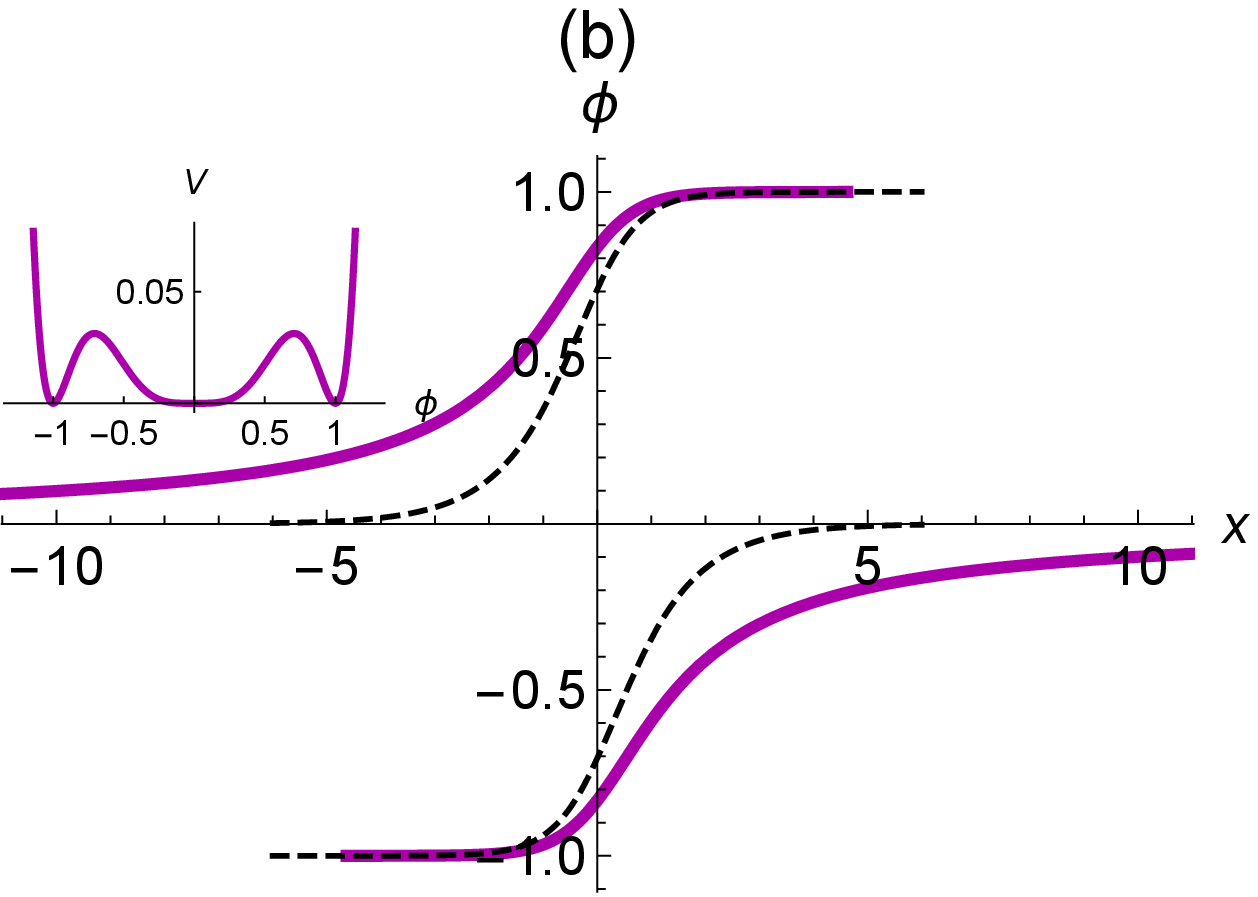}
\caption{Kink solutions of octic field theories with power-law tail asymptotics. (a) The kink from Eq.~\eqref{7.8} ($a=1$ and $\lambda=1/\sqrt{2}$) and the corresponding double well $\phi^8$ potential as an inset. To illustrate the slow tail decay, the $\phi^4$ kink from Eq.~\eqref{eq:V4_kink} is superimposed as a dashed curve. (b) The half-kink from Eq.~\eqref{7.2} ($a=1$ and $\lambda=1$) and the corresponding triple well $\phi^8$ potential as an inset. To illustrate the slow tail decay, the $\phi^6$ half-kinks from Eq.~\eqref{eq:V6_hkink} are superimposed as dashed curves.}
\label{fig:phi8_kinks}
\end{figure}

The tail asymptotics highlighted by Eqs.~\eqref{eq:V8_kink_tail_2dm} and \eqref{eq:V8_kink_tail_3dm}  (illustrated in Fig.~\ref{fig:phi8_kinks})  are in \emph{stark} contrast to the exponentially decaying kinks and half-kinks of the $\phi^4$ and $\phi^6$ models, respectively. Of course, these are not the only examples of double and triple well $\phi^8$ potentials. Other cases are discussed in \cite[Sec.~II]{kcs}, including kink solutions with the ``usual'' exponential tail asymptotics. Furthermore, the slow (algebraic) decay of the tails is indicative of \emph{long-range} interactions of kinks \cite{gani17}. It is noteworthy, that kinks with algebraic tail asymptotics can also be obtained in certain sextic potentials \cite{Bazeia18,Gomes.PRD.2012}. Some initial forays into the excitation spectra of kinks with power-law tails (i.e., linearization about a kink, along the lines of Sec.~\ref{sec:V6_linearize}) were presented in \cite{GaLeLiconf,GaLeLi}.

\subsection{Collisional dynamics and interactions of $\phi^8$ kinks}

\jcmindex{\myidxeffect{C}!Collisions of kinks}
\jcmindex{\myidxeffect{P}!Power-law tails of kinks}

Very little is known about kink collisions under the $\phi^8$ (or any higher-order) polynomial field theory, beyond some preliminary results \cite{Belendryasova.JPCS,Belendryasova.arXiv.08.2017}. The main challenge in studying such collisions is that an ansatz of superimposed single-kink solutions must be used as initial conditions. Thus, while cases of kinks with exponential decay may be studied along the same lines as $\phi^4$ and $\phi^6$ theories (see {Chapter 2} and also recall the discussion and references above), the case of power-law tails is not so simple. In particular, due to the slow algebraic decay of power-law tails, it is neither clear how to quantify the condition of initially ``well separated'' kinks, nor how to decide the truncation length of the finite computational domain. 

For example, even though a $\phi^8$ kink--antikink pair appears to show a weakly \emph{repulsive} character under certain discretizations, resonance windows typical of \emph{attractive} interactions are observed \cite{Belendryasova.JPCS,Belendryasova.arXiv.08.2017}. At this time, this counterintuitive result remains poorly understood, and it is not known how the numerical discretization of the slowly decaying tails affects it. Further mysteries (specifically, unexplained quantitative discrepancies) arise when comparing Manton's \cite{MantonSut,manton_npb} method for estimating the kink--antikink force of interaction to results from the collective-coordinate approach (see \cite{gani17} wherein the kink--antikink force of interaction was  estimated to decay as the fourth power of their separation). The issue of how to numerically discretize kinks with power-law tails, and how to quantify whether they are ``well separated'' initially, is equally thorny \cite{longrange} under the collective-coordinate approach.

\jcmindex{\myidxeffect{C}!Collective-coordinate approach}
\jcmindex{\myidxeffect{M}!Manton's method}

Our current understanding of this subject is evolving. Recent developments suggest that direct numerical simulation approaches that prepare an initial condition for kink--antikink collisions via ``standard'' superpositions (summation or product) of kinks and antikinks do not accurately account for ``long'' (algebraically) decaying tails. As a result, a number of unexpected and, to some degree, unwarranted results arise from collision simulations based on such ans\"atze. To uncover the key physics of kink--antikink collisions in the presence of long-range interactions (power-law tails), the first step is, thus, to determine the proper superposition to be employed in constructing initial conditions. This topic is the subject of ongoing research.

\subsection{Statistical mechanics of the $\phi^8$ field theory and phonons}

\jcmindex{\myidxeffect{P}!Phonon modes}
\jcmindex{\myidxeffect{S}!Statistical mechanics of kinks}

Field theories of the $\phi^8$ type are \emph{not} QES so their statistical mechanics can only be studied by Langevin simulations \cite{AH93,Kovner,BHL99,HL00} or the ``double-Gaussian'' variational approximation \cite{AH93,Kovner}. In principle, one can obtain the lowest energy state numerically, e.g., by Langevin dynamics and use it to calculate the PDF and the concordant thermodynamic quantities. Likewise, the eigenvalues of Eq.~\eqref{eq:schro_evp} can be computed numerically and used in the transfer operator approach. Finally, there exist special cases of the $\phi^8$ field theories with two and three degenerate minima have $V''(\phi_0) = 0$ \cite[Table~I]{kcs}, again leading to the possibility of nonlinear phonon modes. The impact of the latter on the field thermodynamics is, as of now, unexplored.

\jcmindex{\myidxeffect{L}!Langevin dynamics}


\section{Beyond}

There is a veritable zoology of (kink and other) exact solutions in higher-order field theories, depending on the potential specified. Here, we make no attempt to systematically classify or organize these solutions as such an endeavor would be a book on its own. Instead, we highlight some (a) interesting and (b) novel aspects of kinks in higher-order field theories ``beyond'' $\phi^8$.

\subsection{Brief overview of the $\phi^{10}$ field theory}
\subsubsection{Successive phase transitions and kink solutions}

\jcmindex{\myidxeffect{S}!Successive phase transitions}

As in Sec.~\ref{sec:phi8_spt}, one can design a specific $\phi^{10}$ potential, in which varying the coefficient of the $\phi^2$ term leads to a succession of \emph{two first-order} phase transitions \cite[Sec.~III]{kcs}; for brevity, we do not include the latter discussion here. From amongst the many features that $\phi^{10}$ kinks can exhibit, we summarize the following from \cite{kcs}: (a) in the case of five degenerate minima, four quarter-kinks of different energy, e.g., a pair connecting $0$ to $+a$ (or $-a$ to $0$) and a pair connecting $+a$ to $+b$ (or $-b$ to $-a$), for some $a$ and $b$, exist; (b) kinks are generally asymmetric; (c) kinks with power-law tails exist, with a variety of decays possible in the case of three degenerate minima.

\subsubsection{Statistical mechanics of the $\phi^{10}$ theory, including QES results}

\jcmindex{\myidxeffect{S}!Statistical mechanics of kinks}

\begin{figure}[t]
\center
\includegraphics[width=0.5\textwidth]{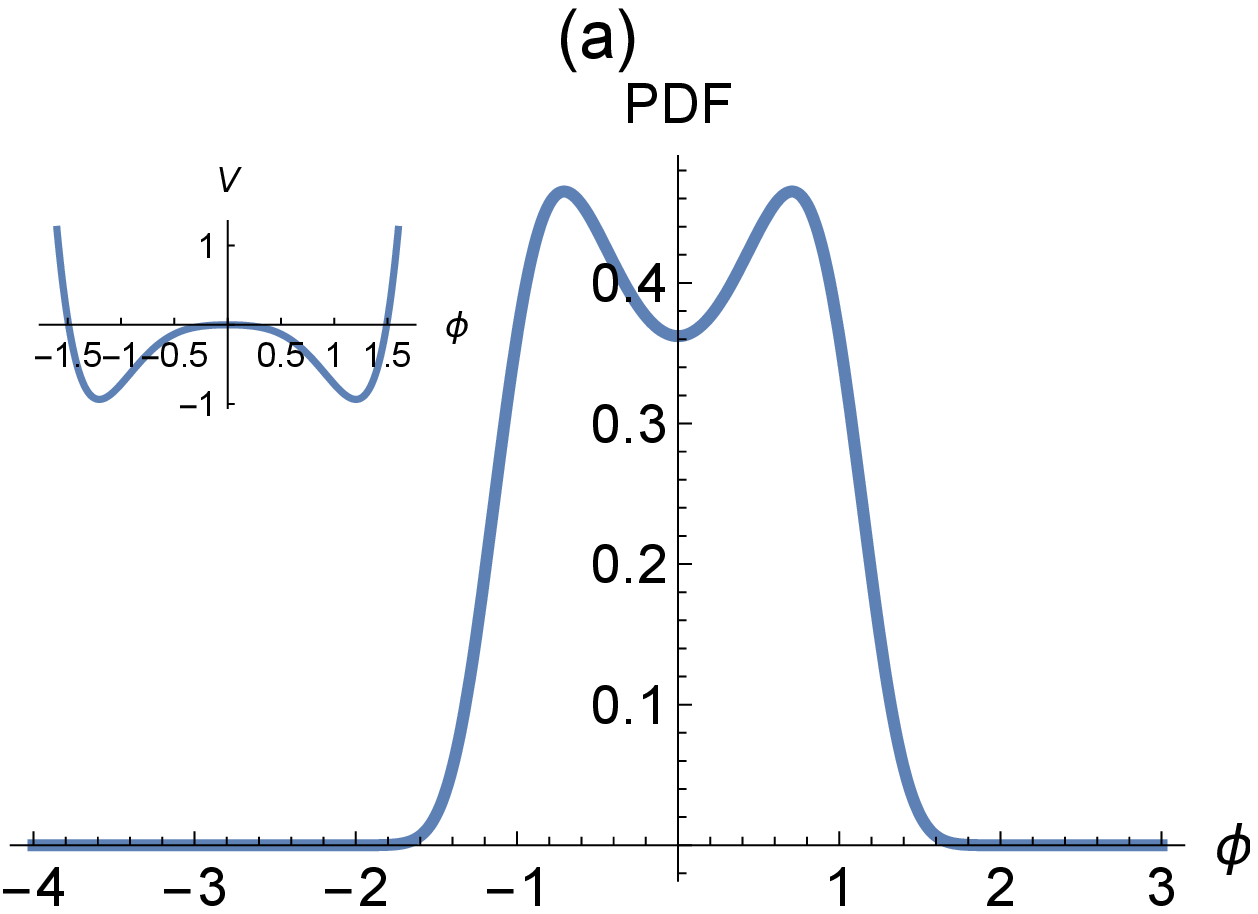}\hfill
\includegraphics[width=0.5\textwidth]{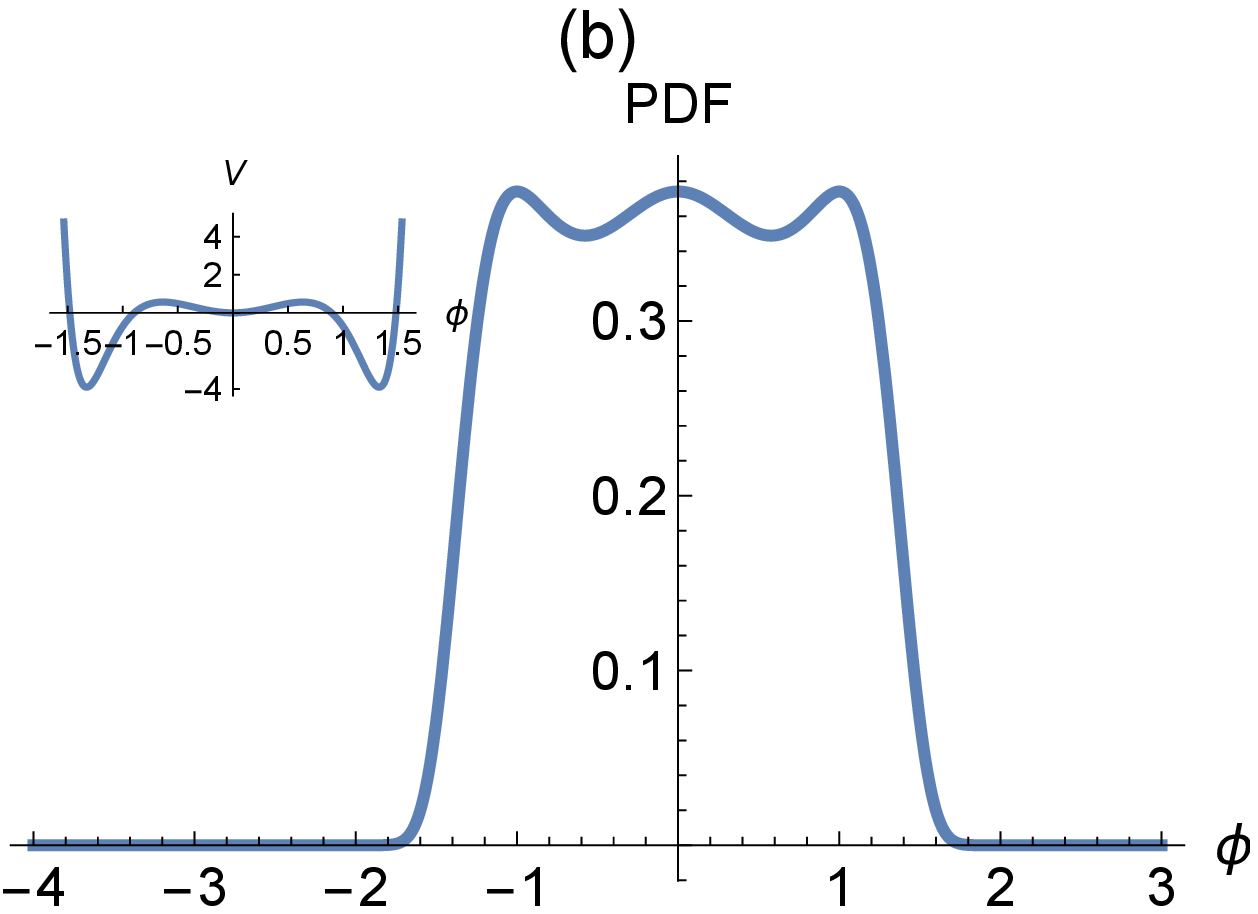}
\caption{(a) Exact PDF for a specific $\phi^6$ potential (i.e., Eq.~\eqref{eq:V6_pt} with $\alpha_2=-1/2$, shown as an inset), based on the wavefunction in Eq.~\eqref{eq:ground_state_phi6} with $K=2$. (b) Exact PDF for a specific $\phi^{10}$ potential (i.e., the generic potential defined by the coefficients $\alpha_{8,6,4,2}$ in Eq.~\eqref{eq:phi10_PDF_cond}, shown as an inset), based on the wavefunction in Eq.~\eqref{eq:ground_state_phi10} with $K=1$.}
\label{fig:compare_6_10_PDF}
\end{figure}

As mentioned above, $\phi^{10}$ is the next example of a QES field theory after $\phi^6$. Following the approach in Sec.~\ref{sec:f6_PDF}, we posit the following generalization of the ansatz in Eq.~\eqref{eq:ground_state_phi6}:
\begin{equation}
\Psi_0(\phi) = \exp\left\{-\frac{\sqrt{2}}{6} \phi^2 \left(\phi^2 - K\right)^2\right\},
\label{eq:ground_state_phi10}
\end{equation}
which clearly has three maxima at $\phi = 0$ and $\phi = \pm \sqrt{K}$, and at these $\phi$ values $\Psi_0(\phi) = 1$, while at all other $\phi$ values $0<\Psi_0(\phi)< 1$. Upon substituting Eq.~\eqref{eq:ground_state_phi10} for the wavefunction $\Psi_0(\phi)$ and a generic tenth-order potential, namely $V(\phi) = \phi^{10} - \alpha_8\phi^8 + \alpha_6 \phi^6 - \alpha_4 \phi^4 + \alpha_2 \phi^2$ as in \cite{kcs}, into Eq.~\eqref{eq:schro_evp} and requiring that equality hold, one obtains two sets of consistency conditions:
\begin{multline}\label{eq:phi10_PDF_cond}
\alpha_8 = \frac{8 K}{3},\quad \alpha_6 = \frac{22 K^2}{9},\quad \alpha_4 = \frac{16 K^3+45 \sqrt{2}}{18},\quad \alpha_2 = \frac{K^4+18 \sqrt{2} K}{9}, \\ E_0 = \frac{K^2}{3 \sqrt{2}}, \quad \beta = 1\,.
\end{multline}
Hence, as long as the potential is of the generic form above, with coefficients $\alpha_{8,6,4,2}$ depending on $K$ as in Eq.~\eqref{eq:phi10_PDF_cond}, then Eq.~\eqref{eq:ground_state_phi10} is an exact ground-state wavefunction (no nodes) of the $\phi^{10}$ field theory at the (inverse) temperature of $\beta=1$ with eigenvalue $E_0 = K^2/(3\sqrt{2})$. The PDF is the normalized squared wavefunction, $\Psi_0^2$. Figure~\ref{fig:compare_6_10_PDF} shows a visual comparison between the exact PDFs obtained herein for the $\phi^6$ and $\phi^{10}$ field theories. Once again, employing different ans\"atze from, e.g., \cite{Ushve} can yield exact excited-state PDFs, as also shown in \cite{kcs}. As discussed above (see also {Chapter 2}), the exactness of these PDFs can be verified via Langevin simulations.

\jcmindex{\myidxeffect{L}!Langevin dynamics}

\subsection{$\phi^{4n+2}$ field theories with three degenerate minima}
\label{sec:phi4n2}

\jcmindex{\myidxeffect{T}!Triple well potential}

Generalizing the result in \cite{CooperPhi2n} on $\phi^{2n+2}$ field theories, let us consider a special family of $\phi^{4n+2}$ field theories with three degenerate minima under a potential of the form
\begin{equation}
V(\phi) = \frac{\lambda^2}{2}\phi^2 (\phi^{2n} - a^2)^2,\quad n=1,2,3,\hdots\,.
\end{equation}
By standard methods, it can be shown that these field theories have \emph{explicit} exact half-kink solutions (connecting $-a^{1/n}$ and $0$ or $0$ and $+a^{1/n}$) given by 
\begin{equation}
\phi_K(x) = \mp \left\{A[1 \mp \tanh(\beta x)]\right\}^{1/(2n)} \,,
\label{eq:phi4np2_kink}
\end{equation}
provided that $A=a^2/2$ and $\beta = \lambda n a^2$. For $a=n=1$, Eq.~\eqref{eq:phi4np2_kink} reduces to Eq.~\eqref{eq:V6_hkink}.

\subsection{Complex, $\mathcal{PT}$-invariant solutions of the $\phi^4$ field theory}
\label{sec:PT_solutions}

\jcmindex{\myidxeffect{P}!$\mathcal{PT}$-symmetry}

Since the introduction of the concept of $\mathcal{PT}$-symmetry in the late 1990s, a host of new physical insights (see, e.g., \cite{BenderRPP} and the references therein) have emerged, resulting in the rapid growth of research on open systems with balanced loss and gain. Here, $\mathcal{P}$ stands for  parity symmetry $\{x,t\} \mapsto \{-x, t\}$, while $\mathcal{T}$ stands for time-reversal symmetry $\{x,t,\mathrm{i}\} \mapsto \{x, -t, -\mathrm{i}\}$. Then, the combined $\mathcal{PT}$-symmetry stands for $\{ x,t,\mathrm{i}\} \mapsto \{-x,  -t, -\mathrm{i}\}$. As before, $\mathrm{i}=\sqrt{-1}$. Recently, novel complex periodic as well as hyperbolic kink solutions with $\mathcal{PT}$-eigenvalue $-1$ have been derived for a number of real nonlinear equations, including the $\phi^4$ and $\phi^{4n+2}$ models, and several higher-order non-polynomial field theories such as sine-Gordon (sG), double-sine-Gordon (DSG) and double-sine-hyperbolic-Gordon (DSHG), etc.~\cite{ks3a,ks3b}. But, while kinks of $\mathcal{PT}$-symmetric nonlinear field theories are not affected by loss/gain, their stability critically depends on the loss/gain profile \cite{Dem13}.

In this section, let us consider a model $\phi^4$ theory: Eq.~\eqref{1.1b} with $V(\phi) = -\frac{a}{2}\phi^2 + \frac{b}{4}\phi^4$. Taking $a,b>0$, $V(\phi)=\mathfrak{C}$ has real solutions $\phi_{1,2,3}$. Then, similarly to how the periodic solutions for $\phi^6$ were constructed in Sec.~\ref{sec:phi6_kink_periodic_etc} (recall Fig.~\ref{fig:V6_kink_lattice}), it can be shown that this $\phi^4$ theory has a kink lattice solution
\begin{equation}\label{4.22}
\phi_{KL}(x) = A\sqrt{m} \sn(\beta x \,|\, m)\,,
\end{equation}
provided that $A = \sqrt{2 \beta^2/b}$ and $\beta = \sqrt{a/(1+m)}$. As before, $\sn$, $\cn$ and $\dn$ are Jacobi's elliptic functions with modulus $m\in[0,1]$ \cite{as}. Equation~\eqref{4.22} reduces to Eq.~\eqref{eq:V4_kink} for $a=b=m=1$.  Remarkably, this same field theory also admits two complex, $\mathcal{PT}$-invariant periodic kink lattice solutions with $\mathcal{PT}$-eigenvalue $-1$:
\begin{subequations}\begin{align}
\phi_{cKL,1}(x) &= A \sqrt{m} \,[\sn(\beta x \,|\, m) \pm \mathrm{i} \cn(\beta x \,|\, m)]\,,\quad \beta = \sqrt{2a/(2-m)}\,,\label{4.24}\\
\phi_{cKL,2}(x) &= A \left[\sqrt{m} \sn(\beta x \,|\, m) \pm \mathrm{i} \dn(\beta x \,|\, m)\right],\quad \beta = \sqrt{2a/(2m-1)}\,,\label{4.26}
\end{align}\end{subequations}
provided that $A = \sqrt{\beta^2/(2b)}$. Notice that, unlike the solution in Eq.~\eqref{4.22}, $a<0$ is allowed here if $m<1/2$ in Eq.~\eqref{4.22}. 

\jcmindex{\myidxeffect{K}!Kink lattice solution}
\jcmindex{\myidxeffect{J}!Jacobi elliptic function}

In the limit $m \to 1^-$, Eqs.~\eqref{4.24} and \eqref{4.26} both reduce to the complex, $\mathcal{PT}$-invariant kink solution
\begin{equation}\label{4.28}
\phi_{cK}(x) = A[\tanh(\beta x) \pm \mathrm{i} \sech(\beta x)]\,,
\end{equation} 
with $A = \sqrt{\beta^2/(2b)}$ and $\beta = \sqrt{2a}$. While the width, $1/\beta$, of the complex, $\mathcal{PT}$-invariant kink in Eq.~\eqref{4.28} is half of the width of the real kink (i.e., Eq.~\eqref{4.22} with $m = 1$), their amplitudes are the same. As described in \cite{cf}, the existence of a complex, $\mathcal{PT}$-invariant kink solution can be traced back to translational invariance: if $\tanh(\beta x)$ is a solution, then so is $\tanh(\beta x + x_0)$. Now, take $x_0 = \mathrm{i}\pi /4$ and observe that $\tanh(\beta x \pm \mathrm{i}\pi/4) = \tanh(2\beta x) \pm \mathrm{i}\sech(2\beta x)$, which immediately substantiates the existence of complex, $\mathcal{PT}$-invariant kink solution with half of the width. Clearly, this argument applies to any model that admits a kink solution of the form $\tanh x$.

\jcmindex{\myidxeffect{C}!Complex kink solution}

Unfortunately, however, a similar argument for the existence of complex, $\mathcal{PT}$-invariant periodic solutions such as those in Eqs.~\eqref{4.24} and \eqref{4.26} is lacking. The obvious generalization would be to argue that if $\sn(\beta x \,|\, m)$ is a solution, then so is $\sn(\beta x+x_0 \,|\, m)$ due to translational invariance. To this end, take $x_0 = \mathrm{i}K'(m)/2$, and, on using the addition theorem for $\sn$, one finds that
\begin{equation}\label{4.31}
\sn\left(\xi \pm \tfrac{1}{2} \mathrm{i} K'(m) \,|\, m\right) = \frac{(1+\sqrt{m})\sn(\xi \,|\, m) \pm \mathrm{i} \cn(\xi \,|\, m)
\dn(\xi \,|\, m)}{m^{1/4} [1+\sqrt{m} \sn^2(\xi \,|\, m)]}\,,
\end{equation}
where, recalling that $K(m)=K'(1-m)$, one has used the fact that (see \cite{as})
\begin{multline}\label{4.32}
\sn\left(\tfrac{1}{2}\mathrm{i}K'(m) \,|\, m\right) = \frac{\mathrm{i}}{m^{1/4}}\,,\quad 
\cn\left(\tfrac{1}{2}\mathrm{i}K'(m) \,|\, m\right) = \frac{\left(1+\sqrt{m}\right)^{1/2}}{m^{1/4}}\,,\\ 
\dn\left(\tfrac{1}{2}\mathrm{i}K'(m) \,|\, m\right) = \left(1+\sqrt{m}\right)^{1/2}\,.
\end{multline}
Inspired by the identity in Eq.~\eqref{4.31}, recently two of us asked \cite{ks4} if there 
is a more general complex, $\mathcal{PT}$-invariant periodic solution. To this end, consider the ansatz 
\begin{equation}\label{4.33}
\phi_{cKL,3}(x) = \frac{A\sn(\beta x \,|\, m) \pm \mathrm{i} B \cn(\beta x \,|\, m)\dn(\beta x \,|\, m)}{1+
D \sn^2(\beta x \,|\, m)}\,,
\end{equation}
where $A$, $B$, $D$ and $\beta$ have to be determined in terms of $a$, $b$ and $m$. After a lengthy calculation, we find that Eq.~\eqref{4.33} is a complex, $\mathcal{PT}$-invariant periodic solution, if
\begin{equation}\label{4.34}
A = \sqrt{(2/b)(D+1)(D+m)\beta^2}\,,\quad \beta = \sqrt{a/(m+1)}\,,\quad  bB^2 = 2D\beta^2\,.
\end{equation}
Unlike the real or the complex periodic kink solutions discussed above, the periodic kink solution in Eq.~\eqref{4.33} exists even if $b < 0$. In particular, if
$-1 < D < -m$, then $b < 0$, which shows that this is a distinct periodic kink solution. In the special case
\begin{equation}\label{4.35}
A = \frac{(1+\sqrt{m})}{m^{1/4}} F\,,\quad B = \frac{F}{m^{1/4}}\,,
\quad D = \sqrt{m}\,,
\end{equation}
the solution in Eq.~\eqref{4.33} takes the form
\begin{equation}\label{4.36}
\phi_{cKL,3}(x) = \frac{F [(1+\sqrt{m})\sn(\beta x \,|\, m) \pm \mathrm{i}\cn(\beta x \,|\, m) 
\dn(\beta x \,|\, m)]}{m^{1/4} [1+\sqrt{m}\sn^2(\beta x \,|\, m)]}\,,
\end{equation}
while the conditions in Eq.~\eqref{4.34} become $F = \sqrt{2m\beta^2/b}$ and $\beta = \sqrt{a/(m+1)}$, which coincide \emph{exactly} with the condition under which the solution in Eq.~\eqref{4.22} exists. 

\jcmindex{\myidxeffect{C}!Complex kink lattice solution}

In the limit $m \to 1$, Eq.~\eqref{4.33} leads to a more general, complex kink solution:
\begin{equation}\label{4.38}
\phi_{cK,3}(x) = \frac{A\tanh(\beta x) \pm \mathrm{i}B \sech^2(\beta x)}{1+D\tanh^2(\beta x)}\,,
\end{equation}
provided that $A = \sqrt{(2/b)(D+1)^2 \beta^2}$, $\beta = \sqrt{a/2}$, and $bB^2 = 2D\beta^2$.  However, Eq.~\eqref{4.38} does not represent a new kink solution. Specifically, as argued above, translational invariance means that given the ``standard'' kink solution $\hat{A} \tanh(\beta x)$, another kinks solution is $\hat{A}\tanh(\beta x+ \mathrm{i}x_0)$. Then, it is easily shown that Eq.~\eqref{4.38} and the standard kink solution $\hat{A} \tanh(\beta x)$ are related via 
\begin{equation}\label{4.39a}
A = (1+D)\hat{A}\,,\quad B = \sqrt{D} \hat{A}\,,\quad D = \frac{1-\cos(2x_0)}{1+\cos(2x_0)}\,. 
\end{equation}
Summarizing, while the complex $\mathcal{PT}$-invariant periodic kink is a new solution, the complex $\mathcal{PT}$-invariant hyperbolic kink is not. Using addition theorems for $\cn$ and $\dn$, more general complex, $\mathcal{PT}$-invariant pulse solutions (similar in structure to Eq.~\eqref{4.33} but with $\mathcal{PT}$-eigenvalue $+1$) have also been recently obtained \cite{ks4}. Determining the stability of these new solutions is an open problem. 

\jcmindex{\myidxeffect{C}!Complex kink lattice solution}

Another observation in the recent work \cite{ks4} is that there exist remarkable connections between the complex solutions of various real scalar field theories. For example, consider a general field under Eq.~\eqref{1.1b} with $V(\phi) = \frac{a}{2}\phi^2 + \frac{b}{2n+2}\phi^{2n+2} + \frac{c}{4n+2}\phi^{4n+2}$, where $n$ is a positive integer. It is amusing to note that, for given $a$, $b$, $c$, and some complex solution $\phi=\phi_1(x)$, then $\phi=\pm \mathrm{i}\phi_1(x)$ is also a solution for the same values of the parameters $a$, $b$, $c$ if $n$ is an even integer (i.e., $n = 2, 4, 6, \hdots$), or with the same $a$ and $c$ \emph{but} with $-b$ if $n$ happens to be an odd integer (i.e., $n = 1, 3, 5, \hdots$). As special cases, $c=0$ and $n=1$ yields a $\phi^4$ field, while $c\ne0$ and $n=1$ yields a $\phi^6$ field. 


\section{Conclusion}

In this chapter, we confined our attention to one-dimensional, higher-than-fourth-order field theories. Just as the $\phi^4$ model has served as a prototype for describing second-order phase transitions and their attendant kinks (or domain walls) as well as breathers, the $\phi^6$ model is a prototype for exploring first-order transitions with a richer phenomenology and different types of kinks. In particular, we discussed exact kink solutions of the $\phi^6$ model, collisional dynamics of various kinks, statistical mechanics of this field theory. However, the $\phi^6$ model is incapable of describing two or more (first- or second-order) successive phase transitions, and we must resort to $\phi^8$ or even higher-order field theories. In this context, we discussed exact kink solutions and their interaction in the $\phi^8$ model and interestingly elucidated the possibility of kinks with power-law tail asymptotics, which is quite different from the exponential tails in the $\phi^4$ and $\phi^6$ field theories. We also briefly considered $\phi^{10}$ as well as a general $\phi^{4n+2}$ field theory with degenerate minima and discussed their kink solutions. Finally, we explored complex, $\mathcal{PT}$-invariant kink solutions of polynomial field theories, and in particular $\phi^4$.
 
\subsection{Open problems}

\jcmindex{\myidxeffect{O}!Open problems}

Beyond the $\phi^6$ model, we merely scratched the surface of the number of open questions for higher-order field theories, their kink-solution collisional dynamics, their statistical mechanics, and their connection to other nonlinear science models, etc. The thermodynamic limiting case of infinite-order (continuous) phase transitions is an exciting area in this vein. The behavior of topological excitations in two (and possibly three) dimensional higher-order field theories is an entirely open issue as well. Nonlocal higher-order field theories, coupled higher-order models and the kink solutions that they harbor remain topics for future investigation.

One of the major open problems in higher-order field theories of the type discussed here is the kink collisional dynamics. Not only is there a far richer phenomenology of kinks in higher-order field theories (including kinks with power-law tails, the difficulties associated with studying their collisions having been mentioned in Sec.~\ref{sec:power-law}), but there are also many more possibilities for pair-wise interaction. The coexistence of kinks with pure power-law (or pure exponential tail) asymptotics with kinks with mixed tail asymptotics (i.e., power-law as $x\to-\infty$ but exponential as $x\to+\infty$) is possible in $\phi^{12}$ field theories with five and four degenerate minima \cite[Secs.~IV-B.2, IV-C.2]{kcs}. What is the nature of these distinct kink-kink interactions? Can we generalize Manton's approach \cite{MantonSut,manton_npb} for calculating kink-(anti)kink effective force of interaction to power-law tails?

\jcmindex{\myidxeffect{C}!Collisions of kinks}
\jcmindex{\myidxeffect{M}!Manton's method}

Furthermore, kinks with different energies can co-exist, as is the case described in \cite[Eqs.~(9)--(17)]{kcs}. To summarize, consider the octic potential with four degenerate minima: $V(\phi) = (\phi^2-a^2)^2(\phi^2-b^2)^2$. It possesses an exact kink solution connecting $-a$ to $+a$ (also found in \cite{Lohe79}) with energy (rest mass) $E_{K,1} = \frac{4\sqrt{2}}{15} a^3(5b^2-a^2)$. There is also an exact half-kink solution connecting $a$ to $b$ (or $-b$ to $-a$) with energy (rest mass) $E_{K,2} = \frac{2\sqrt{2} }{15} (b-a)^3 (b^2+3ab+a^2)$. In \cite{kcs}, it was shown that  
$E_{K,1} \gtreqqless E_{K,2}$ if $b/a \lesseqqgtr 2/(3-\sqrt{5})$. In particular, for $b/a =  2/(3-\sqrt{5})$, the kinks and half-kinks have equal energies. So far, in lower-order field theories, the kinks (and anti-kinks) being scattered necessarily have the same energy because the field theory can have only two ($\phi^4$ and $\phi^6$) or three ($\phi^6$) degenerate minima. Here, for the first time, two kinks of the same type as well as two kinks of different types can exist having equal or unequal energies. Thus, the question to be addressed is: how does the ratio $b/a$ affect kink scattering dynamics? A similar situation occurs in the $\phi^{12}$ field theory with six  degenerate minima \cite[Eq.~(113)]{kcs}.

\jcmindex{\myidxeffect{I}!Internal mode}

Another branch of open questions relates to the fact that the $\phi^{4n+2}$ field theories with three degenerate minima mentioned in Sec.~\ref{sec:phi4n2} offer a \emph{parametrized} way to ``turn up'' the order of the field theory while maintaining the basic exact half-kink structure in Eq.~\eqref{eq:phi4np2_kink}. Thus, we ask: how does $n$ affect kink scattering dynamics, starting with the known case of the $\phi^6$ field theory \cite{dorey} for $n=1$? Furthermore, what is the number of internal modes that the half-kink structure in Eq.~\eqref{eq:phi4np2_kink} possess, and does this number depend on $n$?

It is also worth investigating the stability of the complex $\mathcal{PT}$-symmetric periodic kink solutions of the $\phi^4$ field theory discussed in Sec.~\ref{sec:PT_solutions}.

Finally, we inquire: Have we understood all the connection between the solutions of non-polynomial field theories like sine-Gordon (sG), double-sine-Gordon (DSG), double-sine-hyperbolic-Gordon (DSHG) and the solutions of higher-order polynomial field theories? The former present infinite-order (both periodic and non-periodic) potentials, and some of the early motivation for studying higher-order field theories originated from truncating infinite-order periodic potentials to obtain polynomial field theories \cite{Lohe79} (see also \cite{bazeia06,bazeia11,bazeia13} and \cite[Sec.~V]{kcs}). Specifically, it would be of interest to find out how the nature of kink interactions in non-polynomial theories differs from the corresponding one under a truncated higher-order field theory.


\section*{Acknowledgments}

I.C.C.\ acknowledges the hospitality of the Center for Nonlinear Studies and the Theoretical Division at Los Alamos National Laboratory (LANL), where the authors' collaboration on higher-order field theory was initiated. We acknowledge the support of the U.S.\ Department of Energy (DOE): LANL is operated by Los Alamos National Security, L.L.C.\ for the National Nuclear Security Administration of the U.S.\ DOE under Contract No.\ DE-AC52-06NA25396. I.C.C.\ also thanks V.A.\ Gani and P.G.\ Kevrekidis for many insightful discussions on kinks, collisions, collective coordinates, Manton's method and $\phi^8$ field theory. A.K.\ is grateful to INSA (Indian National Science Academy) for the award of INSA Senior Scientist position.

%


\printindex

\end{document}